%% file: SpitzerLAEpaper_v6.tex
\documentclass[preprint]{emulateapj}
\usepackage{psfig}
\usepackage{graphicx}
\usepackage{apjfonts}
\usepackage{subfigure}
\usepackage{epsfig}
\usepackage{lscape}
\slugcomment{{\sc Accepted to ApJ:} September 14, 2015}

\newcommand{\ang}{$\mbox{\AA}$}
\newcommand{\sol}{$_{\odot}$}

\newcommand{\lya}{Ly$\alpha$}

\def\arcs{\hbox{$^{\prime\prime}$}}
\def\arcm{\hbox{$^{\prime}$}}

\shorttitle{{\it Spitzer} Observations of $z=$ 4.5 LAEs}
\shortauthors{FINKELSTEIN, K.D. ET AL.}

\begin{document}
\title{Probing the Physical Properties of $z=$ 4.5 Lyman Alpha Emitters with {\it Spitzer}}

\author{Keely D. Finkelstein\altaffilmark{1}, Steven L. Finkelstein\altaffilmark{1}, Vithal Tilvi\altaffilmark{2}, Sangeeta Malhotra\altaffilmark{3}, James E. Rhoads\altaffilmark{3}, Norman A. Grogin\altaffilmark{4}, Norbert Pirzkal\altaffilmark{4}, Arjun Dey\altaffilmark{5}\altaffilmark{,6}, Buell T. Jannuzi\altaffilmark{5}, Bahram Mobasher\altaffilmark{7}, Sabrina Pakzad\altaffilmark{8}, Brett Salmon\altaffilmark{2}, and Junxian Wang\altaffilmark{9}}

\begin{abstract}
We present the results from a stellar population modeling analysis of a sample of 162 $z=$ 4.5, and 14 $z =$ 5.7 Lyman alpha emitting galaxies (LAEs) in the Bo$\ddot{\textrm{o}}$tes field, using deep {\it Spitzer}/IRAC data at 3.6 and 4.5 $\mu$m from the {\it Spitzer} Lyman Alpha Survey, along with {\it Hubble Space Telescope} NICMOS and WFC3 imaging at 1.1 and 1.6 $\mu$m for a subset of the LAEs.  This represents one of the largest samples of high-redshift LAEs imaged with {\it Spitzer} IRAC.   We find that 30/162 (19\%) of the $z =$ 4.5 LAEs and 9/14 (64\%) of the $z =$ 5.7 LAEs are detected at $\geq$3$\sigma$ in at least one IRAC band.  Individual $z =$ 4.5 IRAC-detected LAEs have a large range of stellar mass, from 5$\times10^{8}$ -- 10$^{11}$ M\sol.  One-third of the IRAC-detected LAEs have older stellar population ages of 100 Myr -- 1 Gyr, while the remainder have ages $\textless$ 100 Myr.  A stacking analysis of IRAC-undetected LAEs shows this population to be primarily low mass (8 -- 20 $\times$ 10$^{8}$ M\sol) and young (64 -- 570 Myr).  We find a correlation between stellar mass and the dust-corrected ultraviolet-based star-formation rate (SFR) similar to that at lower redshifts, in that higher mass galaxies exhibit higher SFRs.  However, the $z =$ 4.5 LAE correlation is elevated 4--5 times in SFR compared to continuum-selected galaxies at similar redshifts.  The exception is the most massive LAEs which have SFRs similar to galaxies at lower redshifts suggesting that they may represent a different population of galaxies than the traditional lower-mass LAEs, perhaps with a different mechanism promoting Ly$\alpha$ photon escape.
\end{abstract}


\altaffiltext{1}{Department of Astronomy, The University of Texas at Austin, 2515 Speedway, Stop C1400, Austin, TX 78712, USA}
\altaffiltext{2}{George P. and Cynthia Woods Mitchell Institute for Fundamental Physics and Astronomy, Department of Physics and Astronomy, Texas A\&M University, College Station, TX 77843-4242, USA} 
\altaffiltext{3}{School of Earth and Space Exploration, Arizona State University, Tempe, AZ 85287, USA}
\altaffiltext{4}{Space Telescope Science Institute, 3700 San Martin Drive, Baltimore, MD 21218, USA}
\altaffiltext{5}{National Optical Astronomy Observatory, 950 North Cherry Ave., Tucson, AZ 85719, USA}
\altaffiltext{6}{Radcliffe Institute for Advanced Study, Byerly Hall, Harvard University, 10 Garden St., Cambridge, MA 02138, USA}
\altaffiltext{7}{Department of Physics and Astronomy, University of California, Riverside, CA 92521, USA}
\altaffiltext{8}{Gemini Observatory, 670 N. A'ohoku Place, Hilo, HI 96720, USA}
\altaffiltext{9}{CAS Key laboratory for Research in Galaxies and Cosmology, Dept. of Astronomy, University of Science and Technology of China, Hefei, Anhui 230026, China}

\section{Introduction}
Lyman alpha emitting galaxies (LAEs) are thought to be among the youngest galaxies at high redshift ($z=$ 3--6), and they may represent the building blocks of more massive galaxies at lower redshifts (e.g. Malhotra \& Rhoads 2002; Gawiser et al.\ 2007; Finkelstein et al.\ 2011; Malhotra et al.\ 2012). It was first proposed by Partridge \& Peebles (1967) that strong \lya\ emission in high redshift galaxies would be a signpost of primitive galaxies in formation.  This is because \lya\ photons will be produced in large amounts in star forming regions, and the first galaxies should be undergoing periods of extreme star formation. Metallicities in these galaxies will likely be much lower, which will produce hotter stellar photospheres, and hence will also produce considerably more ionizing radiation per unit star formation rate.  Additionally, these early galaxies would contain little dust, which also helps with the escape of resonantly scattered Ly$\alpha$ photons, as they can have long path lengths through the interstellar medium, and thus may have a high probability of being attenuated by dust when present. 

Many studies have been conducted to search for LAEs at high redshift (e.g. Rhoads et al.\ 2000, 2004; Malhotra \& Rhoads 2002; Cowie \& Hu 1998; Hu et al.\ 1998, 2002, 2004; Pentericci et al.\ 2000; Ouchi et al.\ 2001, 2003; Nilsson et al.\ 2007).  Initial studies confirmed that the majority of LAEs appear to be young, low mass galaxies (e.g. Gawiser et al.\ 2006; Finkelstein et al.\ 2007; Lai et al.\ 2007). This makes the study of LAEs very important as it is these low-mass galaxies that are contributing the most to the total star formation activity at these redshifts (e.g. Bouwens et al.\ 2007; Reddy et al.\ 2008). Yet, there is also a population of LAEs that contain some dust, and are not primordial in nature (e.g. Pirzkal et al.\ 2007; Finkelstein et al.\ 2008, Pentericci et al.\ 2009). So, LAEs appear to be a heterogeneous population, with a fraction of them being more evolved and massive.  

To constrain the stellar masses, ages, and dust content of LAEs it is common to perform spectral energy distribution (SED) fitting, comparing photometric observations to stellar population models.  In some of the first samples of detected LAEs, at redshifts 3--6, the ground-based photometry often had to be stacked in order to perform SED fitting (e.g. Finkelstein et al.\ 2007; Nilsson et al.\ 2007; Gawiser et al.\ 2006), providing only average properties of LAEs.  More recent work includes deep {\it Spitzer} IRAC data, which probes the rest-frame optical light for these galaxies, has allowed detailed study of properties of LAEs at high redshifts.  Observations of LAEs with {\it Spitzer} were done at $z=$ 3.1 for a stacked sample of 162 LAEs (Lai et al.\ 2008), where the data were stacked into an IRAC detected and IRAC undetected sample. They find that the IRAC detected sample has an average mass of 9$\times$10$^{9}$ M\sol, and the undetected sample an average mass of 3$\times$10$^{8}$ M\sol, with both stacks best-fit with zero dust.  More recently, results from a stacking analysis by Acquaviva et al.\ (2012) have shown that the LAEs at $z=$ 3.1 were actually best-fit with an older stellar population ($\sim$ 1 Gyr) than LAEs at $z=$ 2.1 ($\sim$ 50 Myr).  This result suggests that these are two very different populations of LAEs, and that the $z=$ 3.1 LAE population cannot evolve directly into the $z=$ 2.1 population. This implication, that the $z=$ 3.1 LAE population is not the progenitor of the $z=$ 2.1 LAE population, may not be that surprising given the typical short lifetime of LAEs. However, this result may also be telling us something about the dangers in estimates from stacking analyses, which makes it difficult to discern any heterogeneity in the population.  For example, Nilsson et al.\ (2011) fit a sample of $z=$ 2.3 LAEs both individually and stacked, and found that while the stellar mass estimates were robust between the stack and the individual objects, the ages and dust attenuations were not.  Vargas et al.\ (2014) examined this in greater detail with a sample of $z =$ 2.1 LAEs with {\it Hubble} photometry, and found that stacking fluxes were able to reproduce the mean properties in a given sample when fit individually, though does not do an adequate job of capturing the large dispersion of LAE properties.

Finkelstein et al.\ (2008, 2009) used deep {\it Hubble} and {\it Spitzer}/IRAC observations of the relatively small GOODS-S field to study 14 LAEs at $z =$ 4.5.  The deep photometry allowed these LAEs to be fit individually, with about 75$\%$ of the 14 galaxies having IRAC detections in at least one band, while the rest had upper limits for all the IRAC fluxes.  The best-fit masses for this sample ranged from 1$\times$10$^{8}$ -- 6$\times$10$^{9}$ M\sol, and with dust ranging from 0.3 -- 5 magnitudes of extinction at 1200 \AA, corresponding to E(B-V) between 0.03 -- 0.4.  At slightly lower redshift, $z=$ 2 -- 3.6,  Hagen et al.\ (2014) used broadband photometry and {\it Spitzer} observations to study a sample of 63 bright LAEs, fitting the objects individually and finding a large range in stellar mass from 3$\times$10$^{7}$ -- 3$\times$10$^{10}$ M\sol. In addition they also found that while most of these bright LAEs had small amounts of extinction, some did have larger amounts of dust, with E(B--V) as large as 0.4.  Thus, LAEs, when examined individually, certainly seem to be a heterogenous population.  To learn more about these intriguing galaxies, especially at higher redshift, we require a larger sample of observed infrared detected LAEs that we can fit individually whenever possible. 

In this paper we present the observations of a large sample of $z=$ 4.5 LAEs detected in {\it Spitzer} IRAC data at 3.6 and 4.5 $\mu$m as well as observed near-IR data at 1.1 and 1.6 $\mu$m with {\it Hubble Space Telescope} ({\it HST}) / NICMOS and WFC3 observations. We perform individual SED fitting on these galaxies using the combination of {\it Spitzer} observations, {\it HST} near-IR data, as well as ground-based optical observations.  Specifically in cases with good {\it Spitzer} IRAC detections, we find the data helps to better constrain the stellar masses and stellar population ages of these galaxies, as it probes the rest-frame, mass-sensitive optical emission, and the age-sensitive 4000 \ang~break. 

In \S 2 we describe the various data sets and observations used in the analysis.  In \S 3 we describe the data reduction steps. In \S 4 we present our results, and describe the SED fitting process and outcomes.  In \S 5 we discuss the implications of our results, comparing to previous studies of LAEs, as well as looking at where these galaxies fall on the main sequence of star formation. In \S 6 we present our summary and conclusions.  Where applicable, we use a cosmology with H$_{\mathrm{0}}$ = 70 km s$^{-1}$ Mpc$^{-1}$, $\Omega_{m}$ = 0.3 and $\Omega_{\lambda}$ = 0.7, and assume a Salpeter (1955) initial mass function (IMF). All magnitudes quoted are in the AB system (Oke \& Gunn 1983).

\section{Observations}
The LAEs targeted by the {\it Spitzer} survey were discovered by the Large Area Lyman Alpha (LALA) Survey (Rhoads et al.\ 2000), which includes the Bo$\ddot{\textrm{o}}$tes field, and has accompanying deep broadband imaging in $B$, $V$, $R$, $I$, and $z^{\prime}$ bands taken with the MOSAIC camera on the 4m Mayall telescope at the Kitt Peak National Observatory. The LAEs in this field have been selected via several narrowband images at different wavelengths ranging from rest-frame H$\alpha$ to $\sim$170 \AA\ red-ward (NB656 H$\alpha$, NB662 H$\alpha$+4,  NB665 H$\alpha$+8, NB670 H$\alpha$+12, NB673 H$\alpha$+16), giving observed LAEs at $z=$ 4.41, 4.45, 4.47, 4.51, and 4.54, respectively. The 5$\sigma$ limiting narrowband magnitudes are 24.8, 24.9, 24.8, 25.1, and 24.2 (AB), in the five narrowband filters respectively, corresponding to a 5$\sigma$ limiting line flux of 2--4 $\times$ 10$^{-17}$ erg s$^{-1}$ cm$^{-2}$.  The seeing in these images was typically 1\arcsec, thus object centroids are known to better than this value.  We also have a few additional sources at $z=$ 5.7, detected with narrowband filters NB815 and NB823 (see Rhoads \& Malhotra 2001 for more detailed description of these observations). To select the $z=$ 4.5 and 5.7 LAE candidates the following criteria were used: (1) a secure detection ($\textgreater$ 5$\sigma$) in the narrowband filter; (2) a strong narrowband excess, i.e.\ the flux density in the narrowband should exceed that in the broadband at the 4$\sigma$ level, this is done by requiring a narrowband -- broadband color $\textless$ -0.75 mag; and (3) no flux at wavelengths shorter than the expected Lyman break. The last condition implies that at $z =$ 4.5, sources are undetected in the B-band, while for $z = $5.7 sources, they are undetected in both the B-band and V-band.

A subset of these narrowband selected LAEs have been observed spectroscopically to confirm the presence of Ly$\alpha$ emission, with the Keck+LRIS or Keck+DEIMOS spectrographs (Dawson et al.\ 2004; Dawson et al.\ 2007).  Of the 162 $z=$ 4.5 LAEs, 48 ($\sim$30$\%$ of our sample) have spectroscopically confirmed Ly$\alpha$ emission at $z\approx$ 4.5.  Based on the spectroscopic follow-up of the larger LALA survey, Dawson et al.\ (2007) estimate a selection reliability of 76$\%$.  LAEs in the LALA field were also followed up spectroscopically with IMACS on the Magellan 6.5m telescope by Wang et al.\ (2009), again finding a spectroscopic success rate of 76$\%$. The other 114 sources in this study have not been targeted spectroscopically and thus are candidate LAEs based on the narrowband and broadband imaging.  Five out of the 14 ($\sim$35$\%$) $z=$ 5.7 LAEs have been spectroscopically confirmed as well.  Table 1 provides a summary of the numbers of LAEs covered at each redshift, those with and without spectroscopic observations.

\input paper_table1.tex

\begin{figure*}[!t]
\vspace{-2cm}
\epsscale{0.9}
\plotone{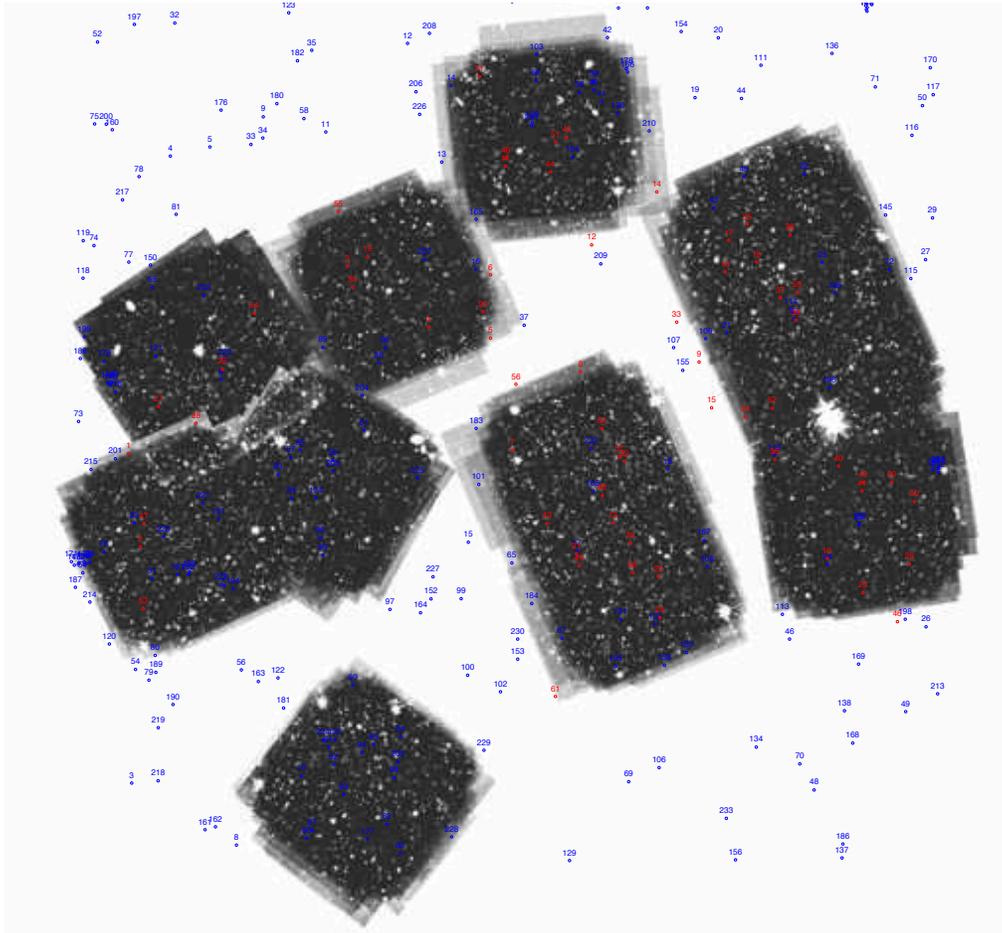}
\vspace{-3cm}
\caption{IRAC 3.6$\mu$m mosaic image of the 4 pointings taken during the {\it Spitzer} Cryo mission plus the additional pointings from the Warm mission, with locations of the candidate LAEs shown in red (spectroscopically confirmed) and blue (candidates based on narrowband imaging).}
\end{figure*}  

The {\it Spitzer} Lyman Alpha survey observed 162 LAEs in the Bo$\ddot{\textrm{o}}$tes Field with deep IRAC 3.6 and 4.5 $\mu$m imaging. These data were taken as part of the Cryo-mission GO cycle 4 and the Warm mission GO cycle 6 (PI Malhotra, PID 40009; PI Rhoads PID 60176).  The {\it Spitzer} IRAC imaging of these objects used 100-second exposure times for each frame, with a total integration time of 13,600 seconds. The original {\it Spitzer} data as part of the Cryo-mission covered approximately 65 LAE candidates. The  second {\it Spitzer} observing program during the Warm mission provided deeper data of the same original pointings along with coverage of another $\sim$ 100 LAE candidates. Additionally we have 800 seconds of {\it Hubble Space Telescope} ({\it HST}) NICMOS and WFC3 imaging at 1.1 and 1.6 $\mu$m covering 27 of the LAEs as part of the ``Physical Nature and Ages of Lyman Alpha Galaxies'' program (PI: S. Malhotra, PID 11153).  The {\it HST} observations directly targeted 23 sources from the spectroscopically confirmed sample only, a few additional non-spec-z sources (narrowband selected objects) happened to be covered in these targeted observations, bringing the total number of LAE sources with {\it HST} NIR coverage up to 27.

\section{Data Reduction \& Analysis}
 The {\it Spitzer} IRAC data were reduced using the MOPEX software (Makovoz \& Khan 2005) provided by the {\it Spitzer} Science Center. The original {\it Spitzer} basic calibrated (BCD) images were drizzled onto a grid of 0.6$^{\prime\prime}$ per pixel. From the combined Cryo mission data we have four pointings; two pointings are approximately 7\arcm$\times$7\arcm~in size, the other two are approximately 7\arcm$\times$12\arcm~in size. The Warm mission data covers the same four pointings, plus five additional 7\arcm$\times$7\arcm~pointings; giving a total coverage area of approximately 510 square arcminutes. Within the MOPEX package we used the standard routines, as well as using the background matching routine, ``overlap.pl'', to match the background between pointings. Figure 1 shows a mosaic of all of the pointings in the 3.6 $\mu$m IRAC band, including both cryo and warm mission data.  

Once the IRAC data were mosaicked together, we used the galaxy fitting software GALFIT (v3.0; Peng et al. 2010) to extract mid-IR fluxes of the {\it Spitzer}-detected LAEs. Due to contamination and crowding in the IRAC images, we used GALFIT to fit and subtract nearby sources around each candidate LAE.  Figure 2 shows three LAE stamps, showing a range in detectable sources and GALFIT modeling and subtraction success for nearby sources. The LAE in the bottom panel is a $z=$ 5.7 candidate LAE, and the middle and top panels are $z=$ 4.5 LAEs.  The three panels for each LAE in Figure 2 show the original IRAC 3.6$\mu$m image, the GALFIT model, and the residual image. For the {\it Spitzer} data, we ran Source Extractor on the IRAC images prior to GALFIT to estimate input magnitudes, radial profile, and positions of other sources in the image.  GALFIT was run on each of the IRAC images on a 30\arcs$\times$30\arcs~region centered on the known LAE location, fitting all sources.  GALFIT requires both an uncertainty image and a PSF.  The uncertainty images from the MOPEX mosaicing of each field were used, and we constructed PSFs using stars in each IRAC mosaic. For galaxies that were detected with GALFIT, we quote the GALFIT flux and error.  For galaxies without IRAC detections, aperture photometry was performed with Source Extractor (Bertin \& Arnouts 1996), using a fixed aperture radius of 9 pixels or 5.4 pixels, on the residual image to obtain a flux measurement at the known source location. In fields where crowding and blending of sources is an issue that was difficult to overcome, even with the careful GALFIT modeling, or where there was a poor GALFIT subtraction of neighboring sources, we used the distribution of residual flux around the LAE to arrive at an independent flux error estimate.  This was done by measuring the surrounding residual flux around the known LAE position, measured with SExtractor using fixed apertures as above, and taking an average residual flux value.  When this value was comparable to the measured aperture flux from the LAE location, then it was set as the flux error, which was deemed more accurate than just taking the flux error from either GALFIT or SExtractor of the LAE itself.

For the {\it HST} WFC3 near-IR data reduction we used the multidrizzle software (2009: Koekemoer et al.\ 2002) to reduce and stack the data with 0.128 arcsec per pixel resolution. The raw images were processed through the CALWF3 task (v2.1: 2010) included in the IRAF STSDAS package, and the  reference files were  obtained from the STScI.  Two of the four images were affected by noticeable persistence due to a bright target exposure prior to these two images. We corrected these two images for persistence using a  persistence removal software for WFC3 IR detectors (K. Long: private communication). In order to align individual images for stacking, we used the 'tweakshift' task to compute these shifts. Finally, the flat-fielded, persistence corrected and spatially matched images are drizzled and combined to produce a final stack.  We followed a similar procedure for both F110 \& F160W filters. For NICMOS data reduction, we used NICRED (Magee et al.\ 2007), a custom-designed pipeline to reduce NICMOS data.   This pipeline  corrects data for persistence, non-linear count rates, bad pixels, and pixels that are affected by cosmic rays. In addition, it performs sky-subtraction before drizzling individual images onto a final mosaic image. The high spatial resolution of the {\it HST} images allowed us to perform aperture photometry to extract the near-IR fluxes.   

The deep ground-based data were reduced using standard IRAF packages, see the reduction details in Rhoads et al.\ 2000.  To measure the ground-based optical broadband and narrowband fluxes, aperture photometry was also performed with Source Extractor, using aperture radii of 4.5 pixels, or 1.2\arcs. Aperture corrections were estimated and applied, based on the difference between the $R$-band Sextractor MagAuto magnitude and the aperture magnitude.

Figures 3--5 show stamps of the 30 $z=$ 4.5 candidate LAEs that are detected at $\ge$ 3$\sigma$ in at least one IRAC band.  Figure 6 shows the 15 LAEs that are only confidently detected in HST NIR bands, but have an upper limit in IRAC ($\textless$ 3$\sigma$).  Figure 7 shows a selection of $z=$ 4.5 and 5.7 LAEs that are not detected in either IRAC or HST, highlighting in some cases the difficulty of extracting IRAC fluxes in crowded / blended regions, or sources where the LAE is too faint and not detected above the IRAC background.
 Figure 8 shows stamps of the $z=$ 5.7 LAEs,  9 of the 14 candidates with IRAC coverage were detected in at least one IRAC band.  Table 1 gives a summary of the total number of sources with IRAC and/or {\it HST} NIR coverage, as well as the number of sources detected in each band at greater than 3$\sigma$ and the resulting detection probabilities.

In order to estimate the range of best-fit stellar properties of the IRAC-undetected sample we performed a stacking analysis.  We performed median flux stacking for all the filters in the IRAC un-detected sample.  This method has been shown by Vargas et al.\ (2014) to be an improvement over median combined image stacking to estimate the fluxes in a stack. We also stacked the IRAC-detected sample in order to compare the average stellar properties between the two populations. To estimate the errors on the median-derived fluxes for each band we performed a bootstrap resampling with replacement of the individual objects in the stack. Stacking was not performed on the $z=$ 5.7 sample, because of the small number of objects in this dataset, as well as the fact that a large majority of this sample was individually detected in IRAC, therefore deeming a stacking analysis unnecessary.

\begin{figure*}[!ht]
\epsscale{0.78}
\vspace{-20mm}
\plotone{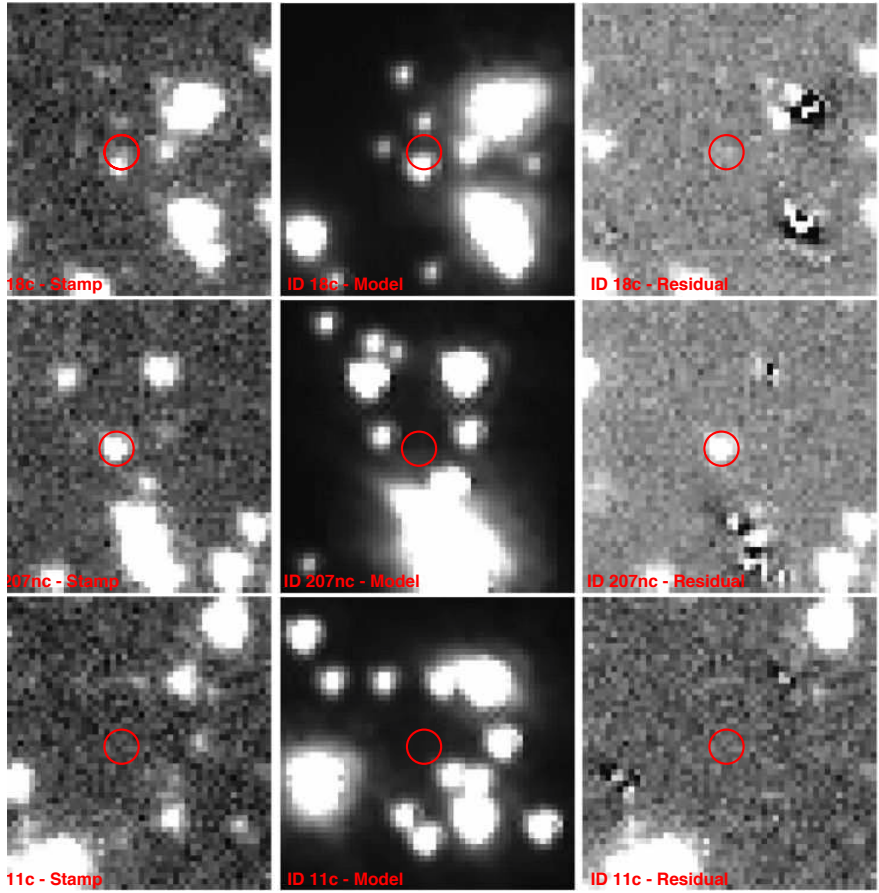}
\caption{Three rows show postage stamps of three LAEs representative of our sample. Left, middle, and right columns show {\it Spitzer} 3.6$]mu$m stamp, GALFIT model, and the residual image, respectively.  The red circle in each panel denotes the position of the LAE based on the ground-based narrowband detection.  Top Row: Object ID 18c, which is a spectroscopically confirmed LAE at $z=$ 4.446 based; the LAE is somewhat hard to detect in the original IRAC ch1 image and is nearby another brighter sources, but becomes easier to perform psf-fitting once the GALFIT models of other sources are subtracted. Middle Row: Object ID 207nc, which is a candidate LAE at $z=$ 5.7, that is easily detected in the original data.   Bottom Row: Object ID 11c, which is a spectroscopically confirmed LAE at $z=$ 4.426, no detections for this object in IRAC were made above the background, either before or after GALFIT psf-fitting was performed.}
\end{figure*}

\subsection{Stellar Population Fitting}
To determine the physical properties of the LAEs we performed spectral energy distribution (SED) fitting, using the updated (2007) version of the Bruzual \& Charlot (2003) stellar population synthesis models. We used all available flux measurements from the $B$-band to the IRAC mid-IR bands (the narrowband fluxes were not used in the fitting).   The observed fluxes for each of the IRAC and/or NIR detected galaxies used in the SED fitting are listed in Table 2 for the $z=$ 4.5 sample and in Table 3 for the $z=$ 5.7 sample.  IRAC flux errors are estimated from GALFIT for the sources with IRAC detections, while flux errors for the aperture-based photometry are taken from SExtractor.  We also apply an additional flux error (5$\%$ of the flux) to all measurements when doing the fitting to account for additional sources of systematic error such as uncertainties in the zeropoints for the ground-based data, and aperture correction differences. 

To estimate the physical properties of each galaxy, we take our flux measurements and compare them to a grid of models, assuming a Salpeter IMF.  The redshift of each source was fixed in the fitting, based on spectroscopic confirmation when available, or otherwise at the redshift that places Ly$\alpha$ in the narrowband filter with the strongest flux excess.  We used empirical intergalactic medium (IGM) opacity (e.g. Madau 1995) in the models, and allowed for varying levels of dust extinction (assuming the extinction curve of Calzetti et al.\ 2000). Our grid parameter space contained 47 discrete values for dust, with E(B-V) between 0 -- 1.  The grid was more finely sampled at small E(B-V) values, and more coarsely sampled at larger values.  We also account for nebular emission by using the prescription of Salmon et al.\ (2015). The nebular emission line strengths depend on the number of ionizing photons which is set by the stellar population age, metallicity, and ionizing escape fraction (see Salmon et al.\ 2015 for more details).  While we do not include the narrowband fluxes as constraints in our SED fitting process, Ly$\alpha$ emission is included in our nebular spectrum, and so the contribution of the observed line to the broadband flux can be accounted for.  However, we acknowledge that the radiative transfer of Ly$\alpha$ is complicated, in that the geometry and kinematics of the interstellar medium (ISM) combined with the resonant scattering properties can result in an emergent Ly$\alpha$ strength which is different than the model predictions (as our model assumes that Ly$\alpha$ experiences the same dust attenuation as continuum photons at similar wavelengths).  Modeling these ISM properties is outside the scope of this analysis, but we note that those studies which have examined the escape of Ly$\alpha$ photons have found that Ly$\alpha$ does not appear to, on average, suffer any ``extra'' attenuation due to these effects (e.g., Finkelstein et al. 2009; Blanc et al.\ 2011).  This is confirmed as the ratio of the observed Ly$\alpha$ flux to that of the best-fit model predicted flux has a median value close to unity (for IRAC-detected sources).  However, there is a non-negligible spread, in that $\sim$40\% of these galaxies have narrowband observed-to-model flux ratios which differ by more than a factor of two from unity.  Thus, while Ly$\alpha$ does not appear to be dramatically attenuated more than continuum photons on average, this can clearly vary on a galaxy by galaxy basis.

Ages of galaxies in the models are allowed to vary from 10 Myr to the age of the universe at these redshifts; with specific grid values in steps of 1 Myr between 10 -- 20 Myr; in steps of 5 Myr between 20 -- 100 Myr; and in steps of 100 Myr between 100 -- 1000 Myr .  Metallicity is also allowed to vary with discrete values of [0.02, 0.2, 0.4, 1.0] Z\sol. Star formation histories in the models are varied, with exponentially decaying and increasing star formation timescales.  SFH = exp$^{-t/\tau}$, with $\tau$ = [0.1, 1, 10, 100, 1000, 10$^{4}$, -300, -1000, -10$^{4}$] Myr.  The shortest (0.1 Myr) and longest (10 Gyr) values of $\tau$ approximate a single burst, and continuous star formation, respectively.  Model fluxes are calculated by integrating the model spectrum weighted by each filter's transmission curve.  We then find the best-fit model via $\chi^{2}$ minimization, with $\chi^{2}$ defined as:

\begin{equation}
\chi^2 = \sum\limits_{filters} \frac{(f_{obs} - f_{mod})^2}{\sigma_{obs}^2}
\end{equation}

We fit the SEDs of individual galaxies with IRAC detections or IRAC upper limit constraints to stellar population synthesis models.  To determine the uncertainty on our best-fit stellar population properties we ran 10$^{3}$ Monte Carlo simulations on each object, varying the input flux measurements by an amount drawn from a Gaussian distribution with a standard deviation equal to the flux errors for a given filter.  We then determine the 68\% confidence range by finding the central 68\% of results for a given parameter.  We list the best-fit and 68$\%$ confidence interval results from the SED fitting for each galaxy (both the $z=$ 4.5 and $z=$ 5.7 samples) in Tables 4 --6.

\section{Results}

\subsubsection{Stellar Mass Ranges -- Results with and without Nebular Emission}
To study the effects that nebular emission, especially H$\alpha$ emission in the 3.6 $\mu$m band, has on the derived stellar mass and other properties, we fit the galaxies both with and without nebular emission. 

Figures 9--14 show samples of the best-fit SEDs for the {\it HST}/NIR and IRAC-detected objects, and are fit including nebular emission. We plot the 1$\sigma$ error bars on each data point, and fluxes with less than a 3$\sigma$ detection are shown by an upper limit. Figures 15 -- 17 plot a sample of the spectroscopically confirmed objects which have detections in the IRAC bands (the same sample that is shown in Figures 8 -- 10), but this time the objects have been fit without including nebular emission (all other allowable SED fitting parameters were kept the same).  For the most part, especially for the IRAC-detected sample, the quality of the fits without nebular emission tend to be worse, with larger $\chi^{2}$ values derived from the best-fits. For the IRAC-detected LAEs fit with nebular emission, we find that the typical masses range from 5$\times10^{8}$ -- few $\times$10$^{11}$ M\sol.  

The ages also show a large range, with about one-third of the IRAC-detected LAEs having stellar population ages between a few 100 Myr -- 1 Gyr, and the other two-thirds with ages between 10 Myr -- 100 Myr.  When these same galaxies are fit without nebular emission, we find the ages are consistently higher, with a majority having best-fit ages between 600 Myr -- 1 Gyr.  When comparing the best-fit stellar population ages of the IRAC detected sample in the two different fitting methods, we find only 11 out of 30 LAEs (37$\%$) have ages greater than 100 Myr when fitting with nebular emission, whereas 22 out of 30 (73\%) have ages over 100 Myr when fitting without nebular emission. We find a mean age ratio of  Age$_{neb}$ / Age$_{no ~neb}$ $=$ 0.5; an individual LAE is 50$\%$ more likely to have a best-fit stellar population age greater than 600 Myr if the fit is done without including nebular emission. Nebular emission is thus a crucial ingredient when fitting this population of star-forming galaxies.

Figure 18 shows a histogram of the best-fit masses for the $z=$ 4.5 LAEs, including errors determined from the Monte Carlo simulations for the IRAC-detected, and the NIR-detected sample and optical-only detected samples with upper limits in one or more of the IRAC bands. The IRAC detected objects are typically more massive than the LAEs that are only detected in the NIR or optical by one order of magnitude. The best-fit masses for both the IRAC and NIR samples tend to have typical mass errors of $\sim$0.2 dex, as compared to 0.4 dex for the IRAC non-detected sample, demonstrating that the addition of the IRAC fluxes leads to more robust SED fitting results, specifically on the stellar mass estimates.  While the {\it HST}-only detected sources do not have full IRAC detections, they all have IRAC upper limits, and our fits show that when we have reliable HST NIR data that is coupled with IRAC upper limits which are close to the {\it HST} detections, it becomes a reasonable constraint on the SED fits.  Typically in these cases, a large 4000 \AA break is not allowed, and we end up with a lower stellar mass sample, with reasonable constraints for the {\it HST}-only detected LAEs.  When the IRAC upper limit is not close to the HST detections, then the fit is less constraining and we do have higher uncertainties on the stellar mass estimates for these LAEs. The best-fit masses for the two stacks (IRAC-detected; IRAC-undetected) are also plotted in Figure 18.  By stacking the individual LAEs we saw improvement in the range of best-fit masses, especially for the IRAC-undetected stack as compared to individual LAEs in the blue histogram.

Figure 19 shows the mass histogram for the $z=$ 5.7 candidate LAEs. Nine objects are detected in IRAC and/or the NIR. These objects are some of the brightest in the IRAC sample, and appear to be relatively massive at this redshift.  There are four objects at this redshift that only have measured upper limits in IRAC, and these tend to be less massive.  

\subsubsection{Age and Dust Results}
The best-fit age and E(B-V) for each galaxy are shown in Figure 20.  The errors on the age distribution and dust are less constrained, and we see a spread in both age from 10 Myr -- 1000 Myr, and dust from 0 -- 1 magnitudes of dust reddening. LAEs with best-fit ages of 10 and 1000 Myr are falling right on the parameter endpoints in our grid, therefore we cannot say anything meaningful about the specific stellar population age in these cases, especially those that have 68\% confidence in age that lies in one grid space, e.g. from 10 -- 10 Myr or 1000 -- 1000 Myr, and we can only conclude that these LAEs are likely very young or old.

As has been shown in Pirzkal et al.\ (2012), as the photometric uncertainty increases in the observations, degeneracy between stellar ages and extinction also increases. The addition of the IRAC data is expected to help break this degeneracy; however when looking at typical uncertainties in both age and dust parameters for both the IRAC-detected sample and IRAC-undetected sample, we do not see any large systematic improvement in the range in age and dust estimates. However, we do see in some of the objects (but not all) with reliable HST NIR data, that they have more robust constraints on the stellar age and dust parameters for these individual galaxies.  A broad conclusion we can make, is that while $z=$ 4.5 LAEs are typically young (ages $\textless$ 100 Myr), there is a smaller population of more evolved LAEs, with ages between 500 - 1000 Myr, and this population tends to be have smaller amounts of dust.  Overall, for the IRAC detected sample, we see a typical uncertainty on the stellar age of $\pm$0.35 dex, compared to $\pm$0.42 dex for the non-IRAC detected sample; similar uncertainties of $\pm$ 0.07 (IRAC detected) and 0.12 (non-IRAC detected) on E(B -- V) are seen for both samples.

\begin{figure*}
\epsscale{0.9}
\plotone{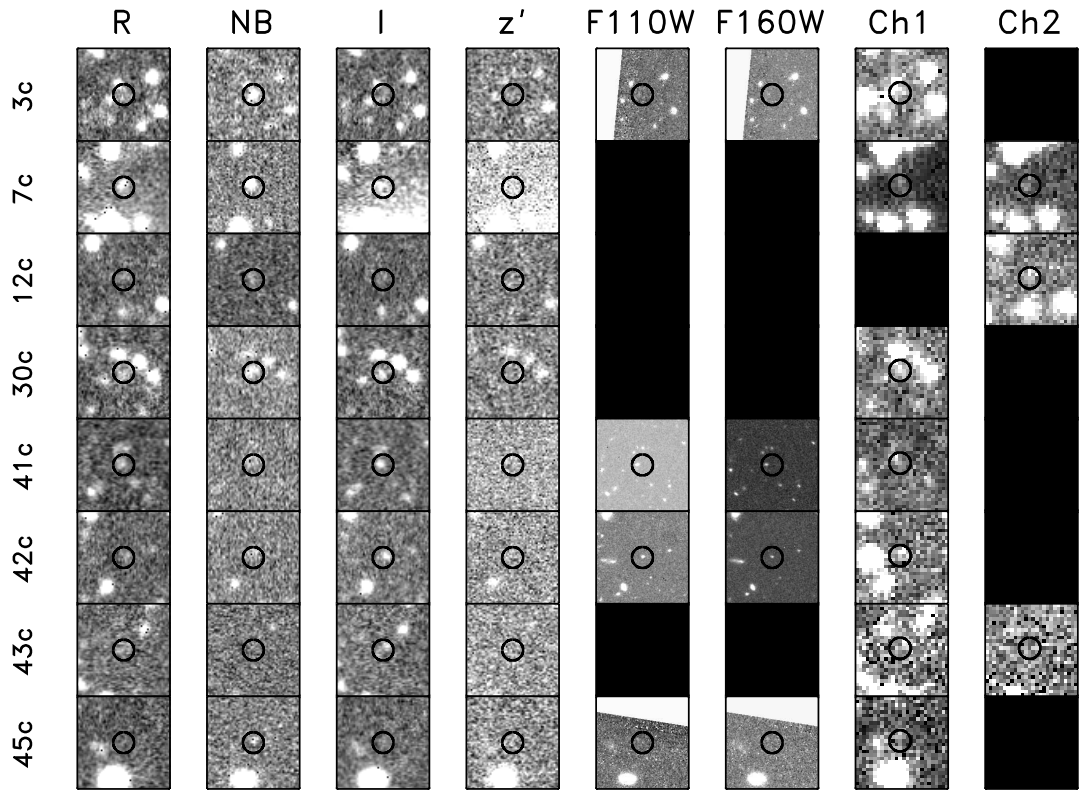}
\caption{Postage stamps of $z=$ 4.5 candidate LAEs in observed wavelengths: broadband R, narrowband H$\alpha$, broadband I, and z', {\it Hubble} NIR imaging at 1.1 and 1.6 $\mu$m, and {\it Spitzer} mid-IR at 3.6 and 4.5 $\mu$m. Stamps are 15\arcs on a side, and the circular apertures have radii of 2\arcs.  All of the LAEs shown in Figures 3--5 are detected at the 3$\sigma$ level (or higher) in at least one IRAC band. The LAE position denoted by the circle is based on the narrowband detection.}
\end{figure*}

\begin{figure*}
\epsscale{0.9}
\plotone{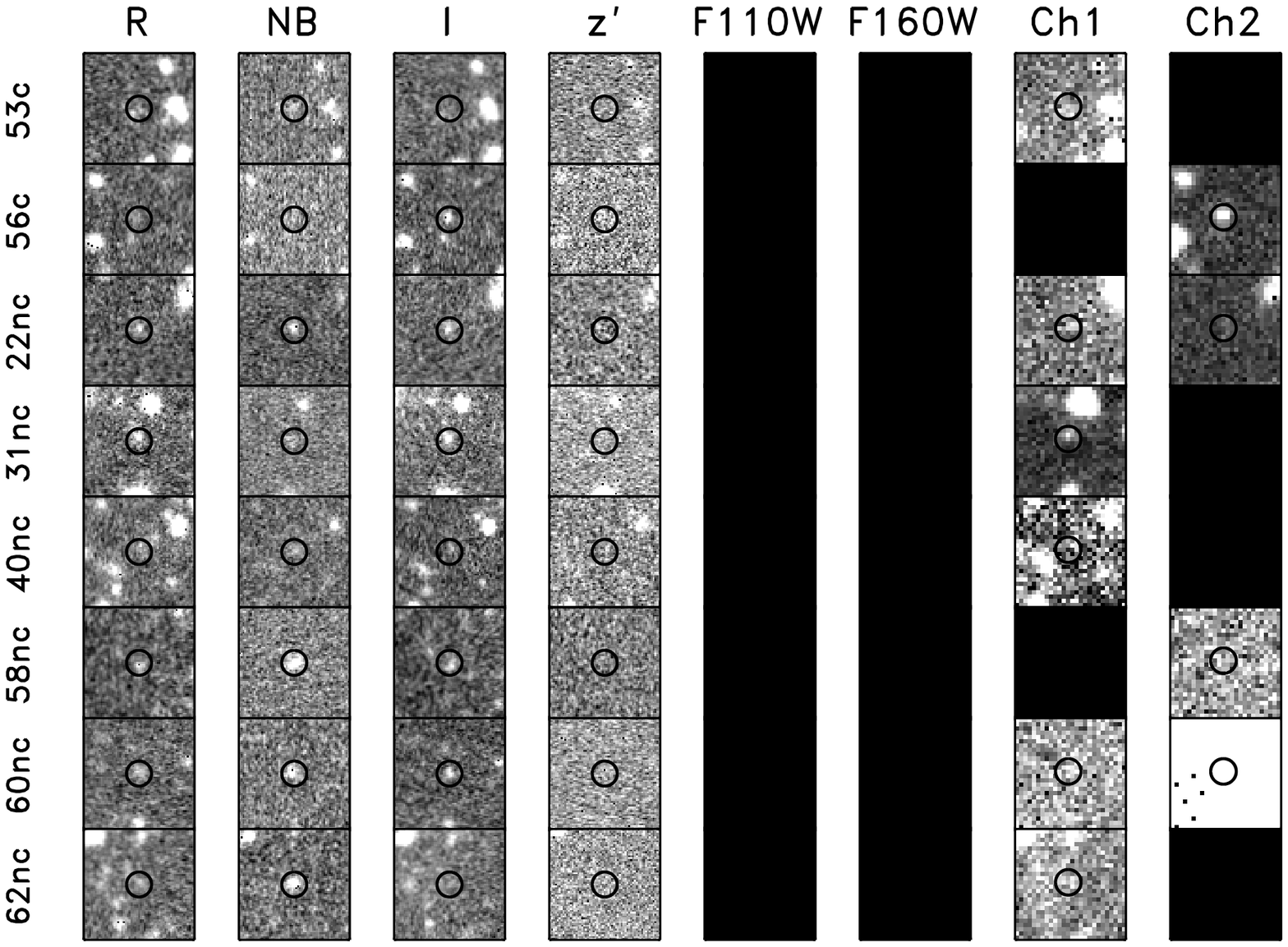}
\vspace{1cm}
\plotone{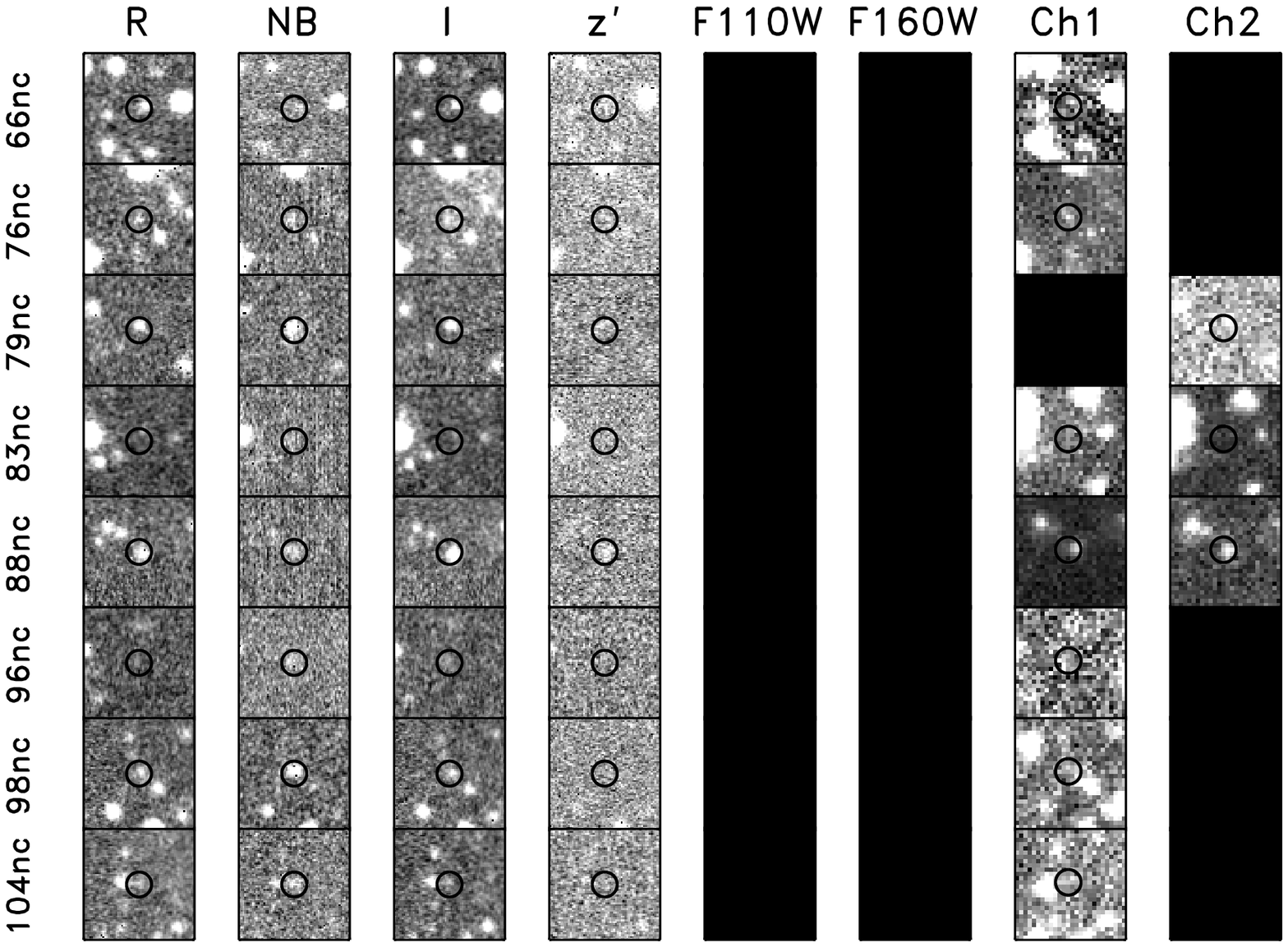}
\caption{Same as Figure 3}
\end{figure*}

\begin{figure*}[!t]
\epsscale{0.9}
\plotone{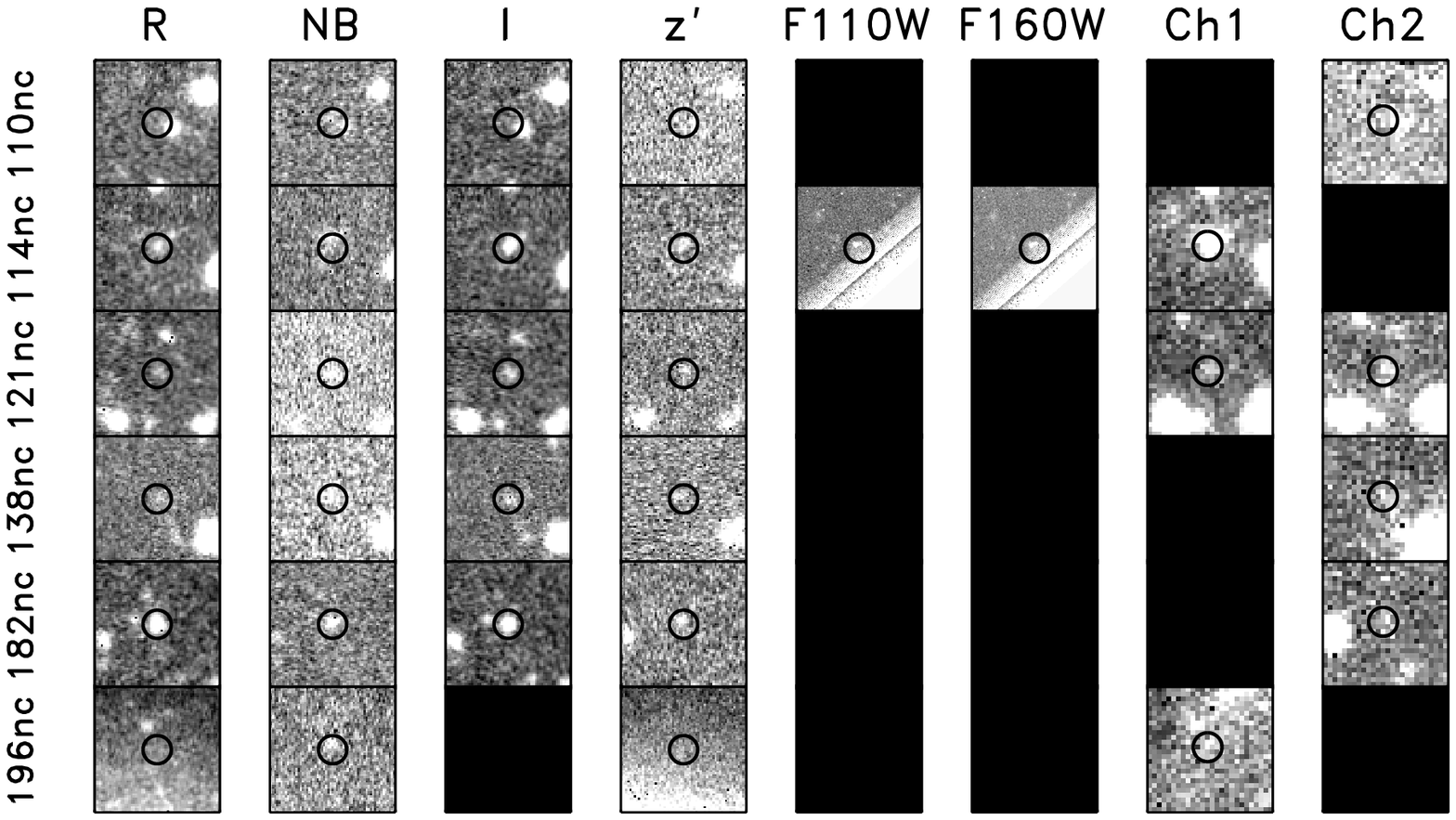}
\vspace{-2cm}
\caption{Same as Figure 3}
\end{figure*}

\clearpage

\begin{figure*}[!t]
\vspace{1cm}
\epsscale{0.9}
\plotone{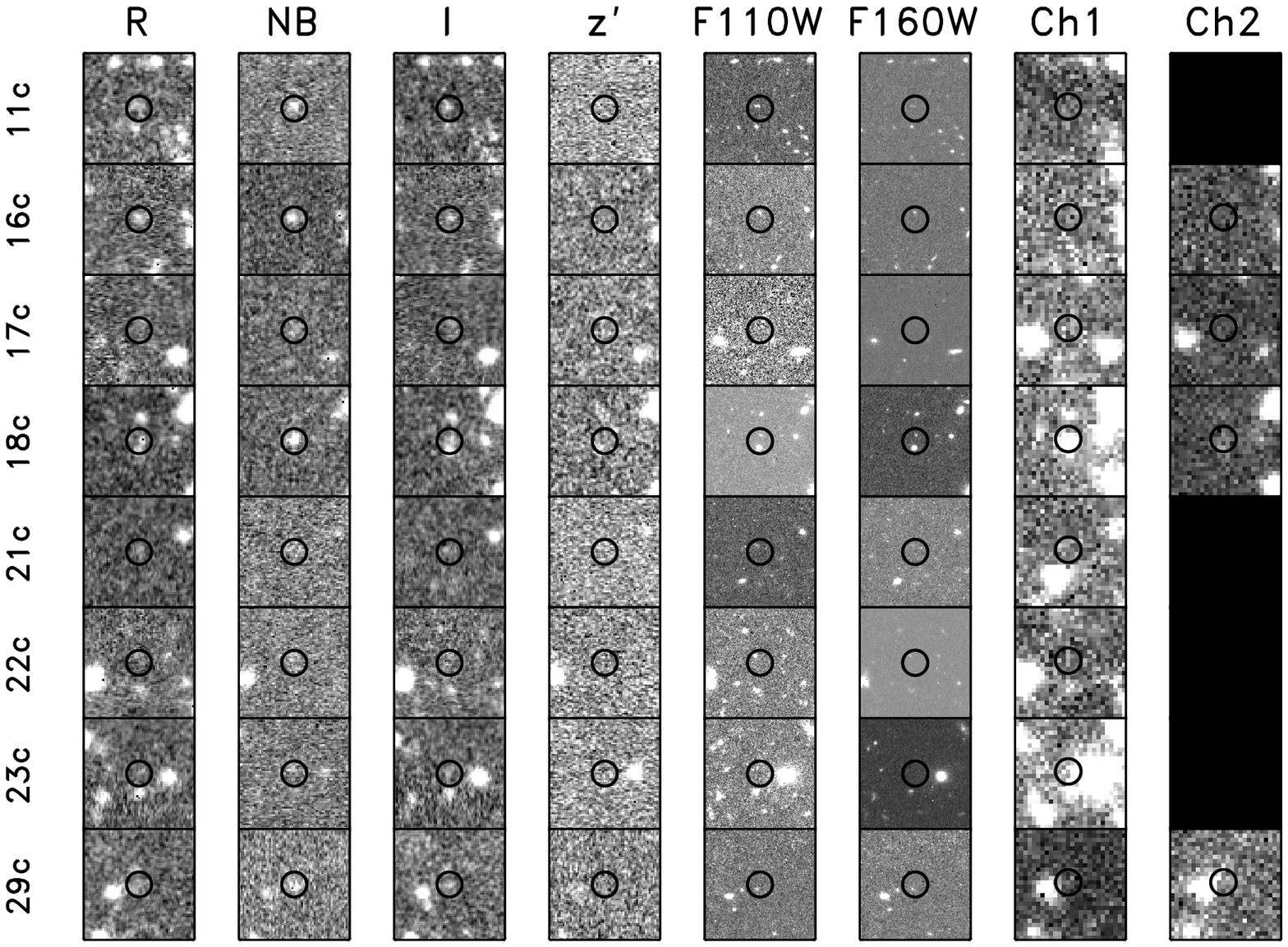}
\vspace{1cm}
\plotone{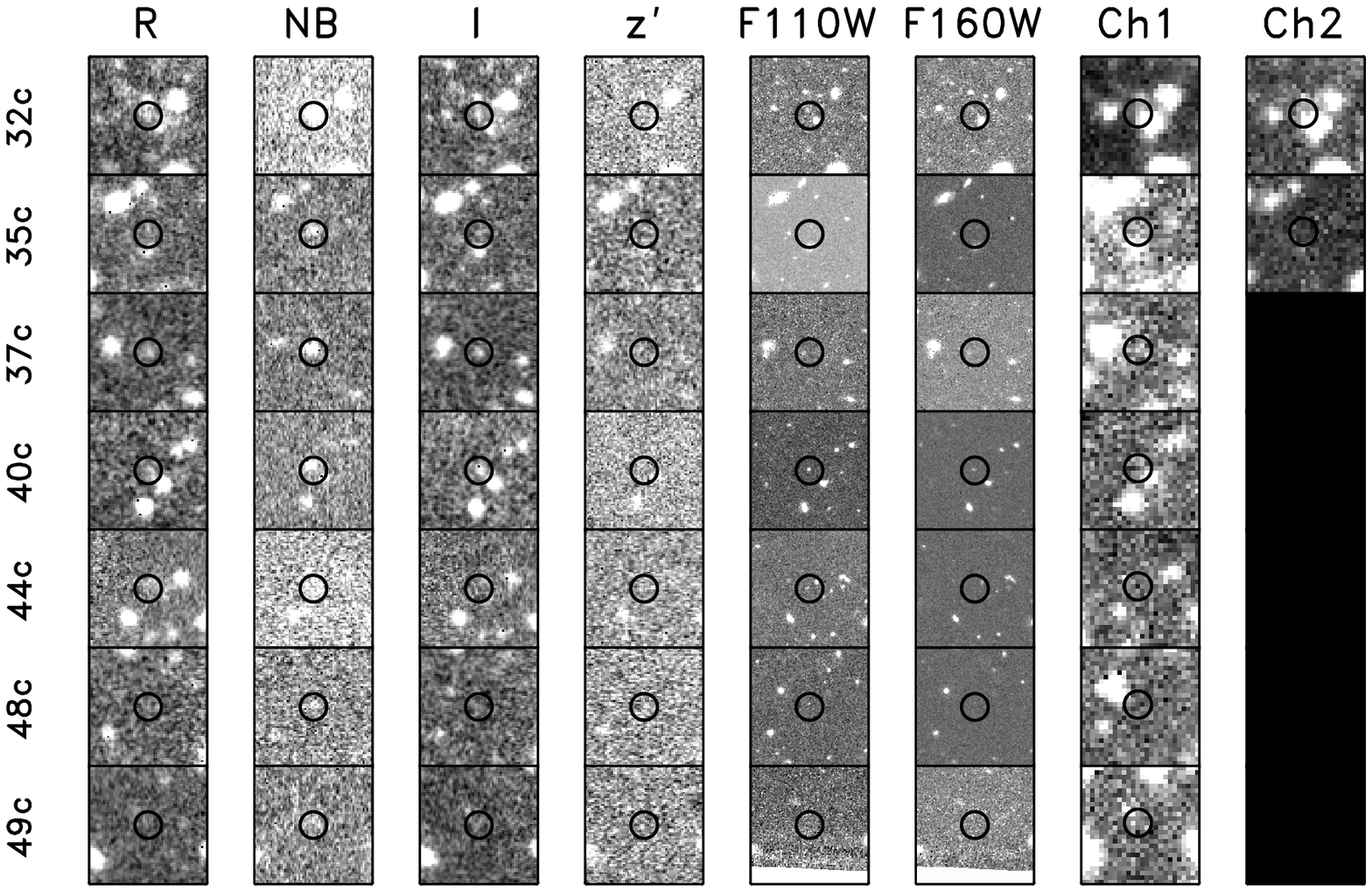}
\vspace{-1cm}
\caption{Same as Figure 3, except that these sources are only reliably detected in HST NIR bands, and have IRAC upper limits.}
\end{figure*}

\begin{figure*}[!t]
\vspace{1cm}
\epsscale{0.9}
\plotone{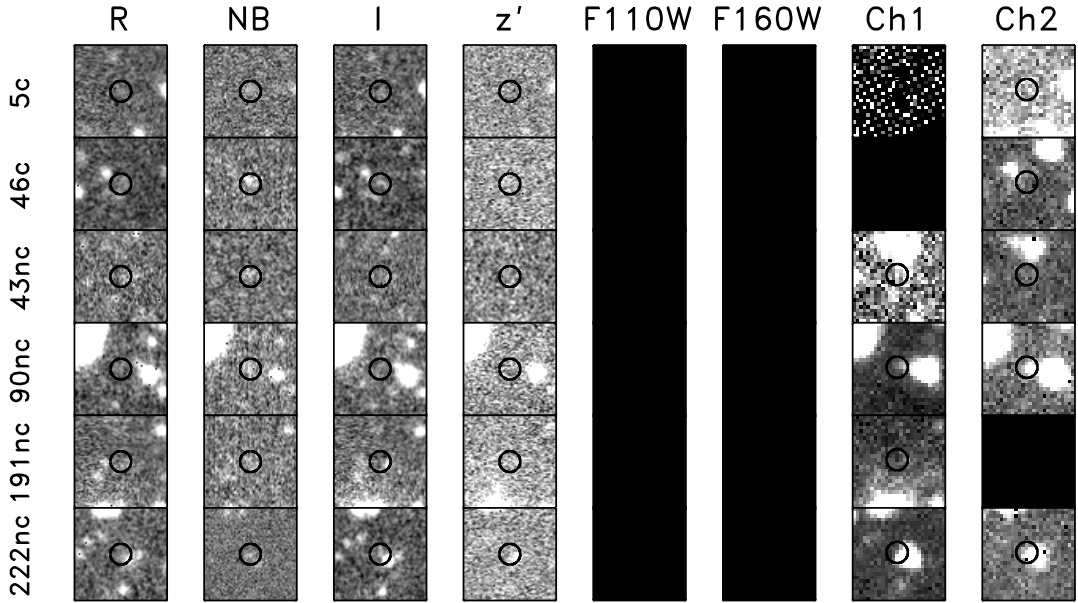}
\vspace{-2cm}
\caption{Same as Figure 3, except that these sources are not reliably detected in IRAC, and only have IRAC upper limits. These sources are examples where the IRAC flux was difficult to measure based on crowding or blending of sources, or cases where the LAE was too faint and not detected at all. Source ID 222 is a $z=$ 5.7 LAE, but it also suffers from crowding issues, and a reliable flux for it could also not be measured.}
\end{figure*}

\begin{figure*}
\vspace{1cm}
\epsscale{0.9}
\plotone{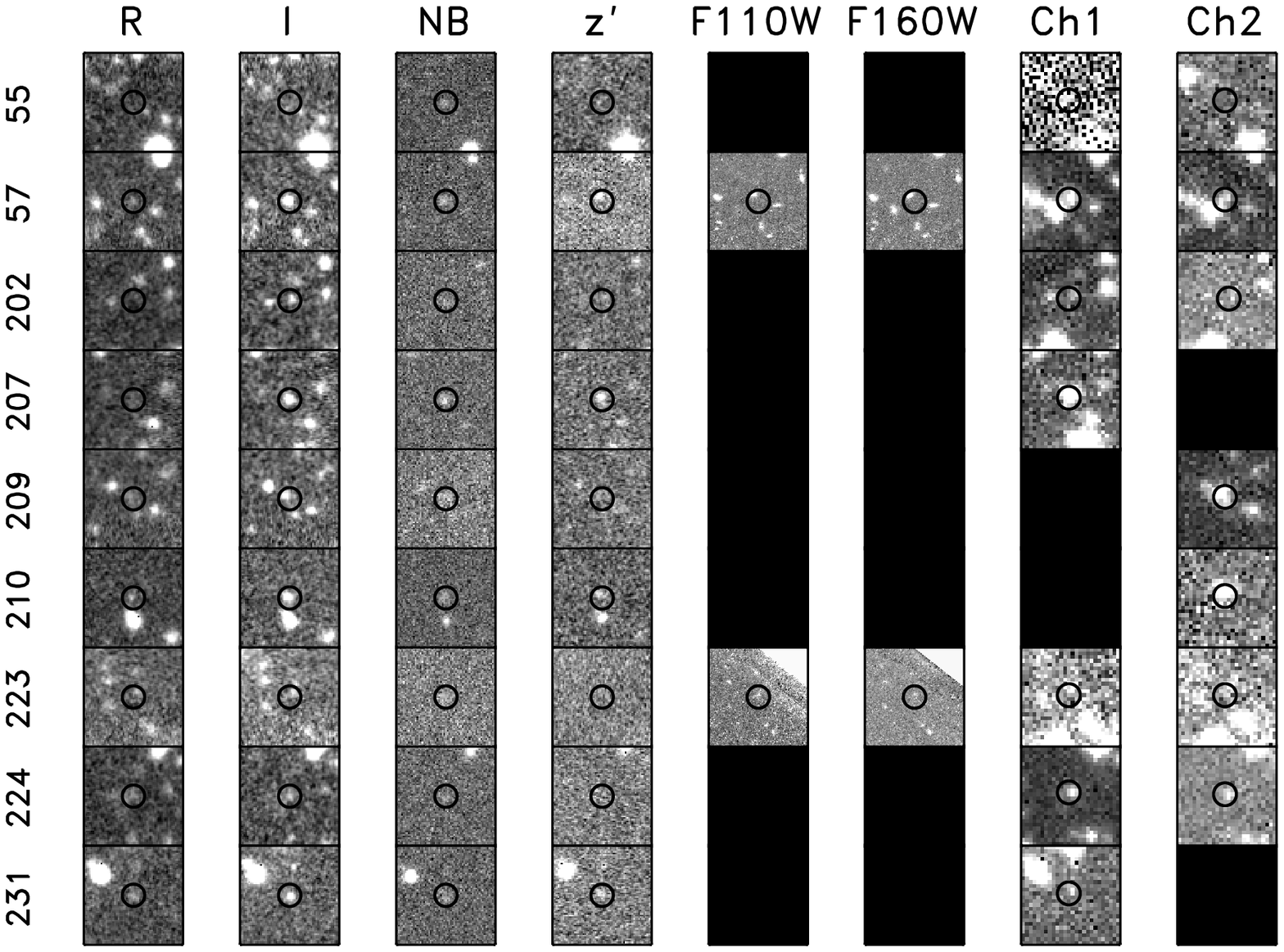}
\caption{Postage stamps of $z=$ 5.7 candidate LAEs in observed wavelengths: broadband R, I, narrowband F815 or F823, z', and {\it Hubble} NIR imaging at 1.1 and 1.6 $\mu$m, and {\it Spitzer} mid-IR at 3.6 and 4.5 $\mu$m.}
\end{figure*}

There is a small trend of increasing age with decreasing dust content, but again the errors on both parameters show that we cannot reliably constrain the stellar ages with the current models and observations, therefore making it difficult to make strong conclusions about this parameter.  We do see that in general the smaller sample of optical and NIR detected objects only (green data points) tend to be young and with only a smaller amount of dust, from 0 -- 0.4 magnitudes of dust.  The more massive sample of IRAC detected LAEs (red points) tend to span a wider range in both properties, as do the optical only / IRAC upper limit detected LAEs. Overall, for both of these samples we see that the classical SED fitting is still not allowing us to break the issue of age / dust degeneracy, and therefore we cannot fully explore this issue.

\begin{figure*}
\epsscale{0.8}
\plotone{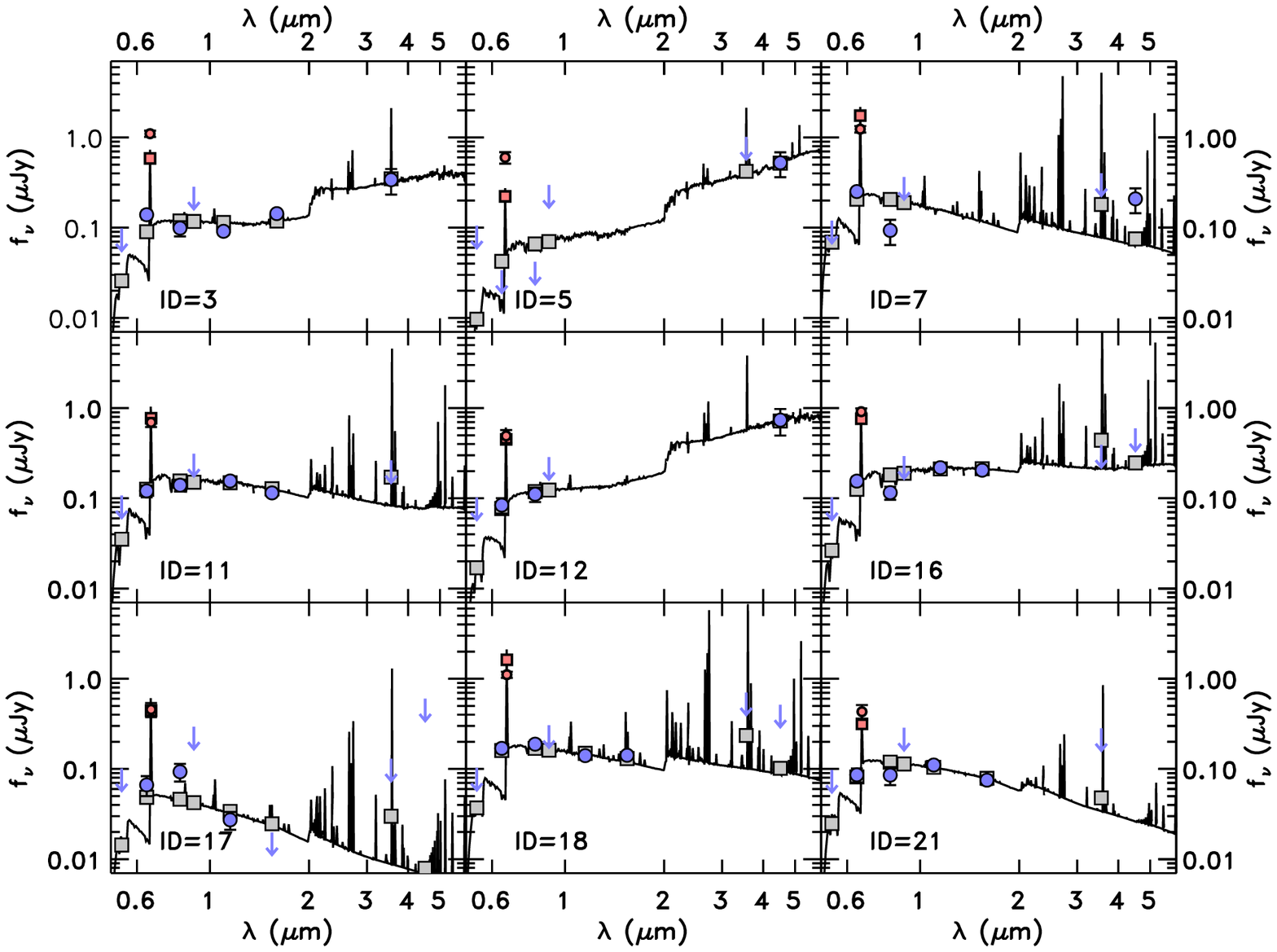}
\vspace{4mm}
\caption{Spectral energy distribution fits for individual $z=$ 4.5 LAEs with IRAC detections and/or HST NIR detections.  These 9 sources shown are from the spectroscopically confirmed sample. Observed fluxes are shown by the blue data points, blue arrows are 2$\sigma$ upper limits.  Red points show the observed narrowband selection flux and best-fit model average flux (circle and square, respectively), however the narrowband flux was not used in the fitting process to constrain the best-fit SED model.  The model average bandpass fluxes for the broadband filters are shown by the grey squares.  When performing the SED fitting we allowed for varying star formation histories, varying amounts of dust, ages, metallicity, and nebular emission. The known redshift from the spectroscopy was used as a constraint in the modeling for these sources.}
\end{figure*}

\begin{figure*}
\epsscale{0.8}
\plotone{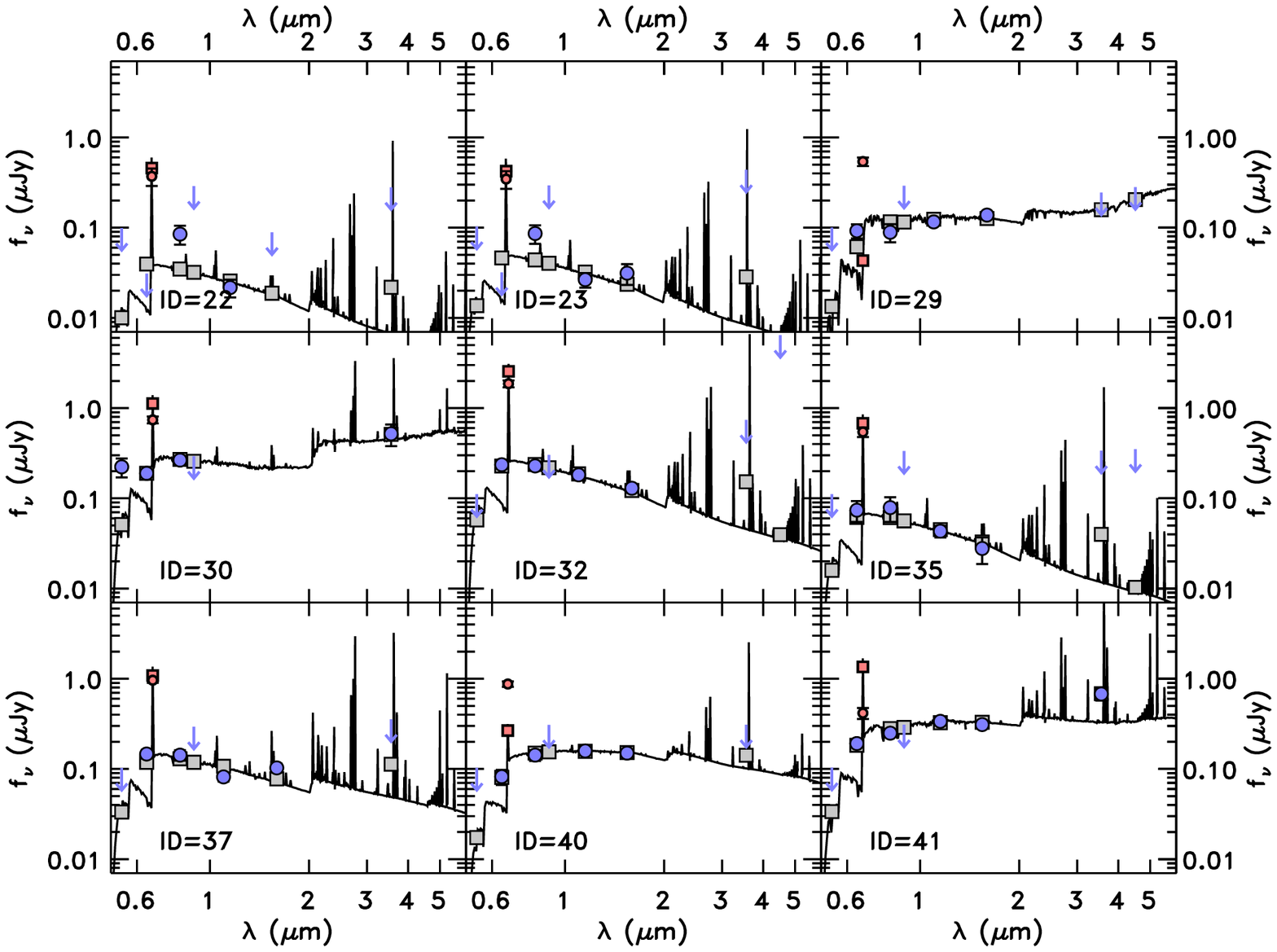}
\vspace{4mm}
\caption{Same as Figure 9}
\end{figure*}

\begin{figure*}
\epsscale{0.8}
\plotone{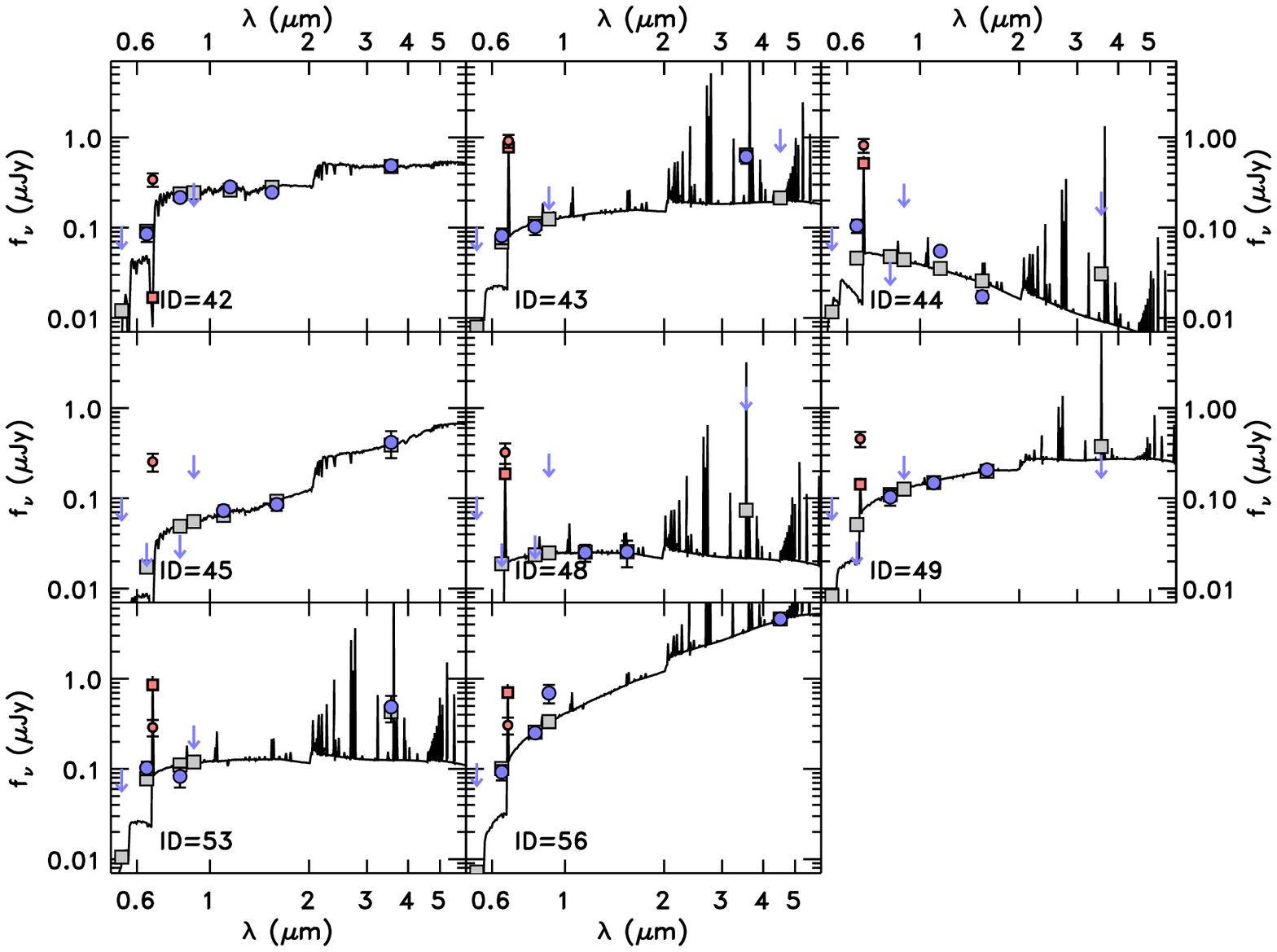}
\vspace{4mm}
\caption{Same as Figure 9}
\end{figure*}

\begin{figure*}
\epsscale{0.8}
\plotone{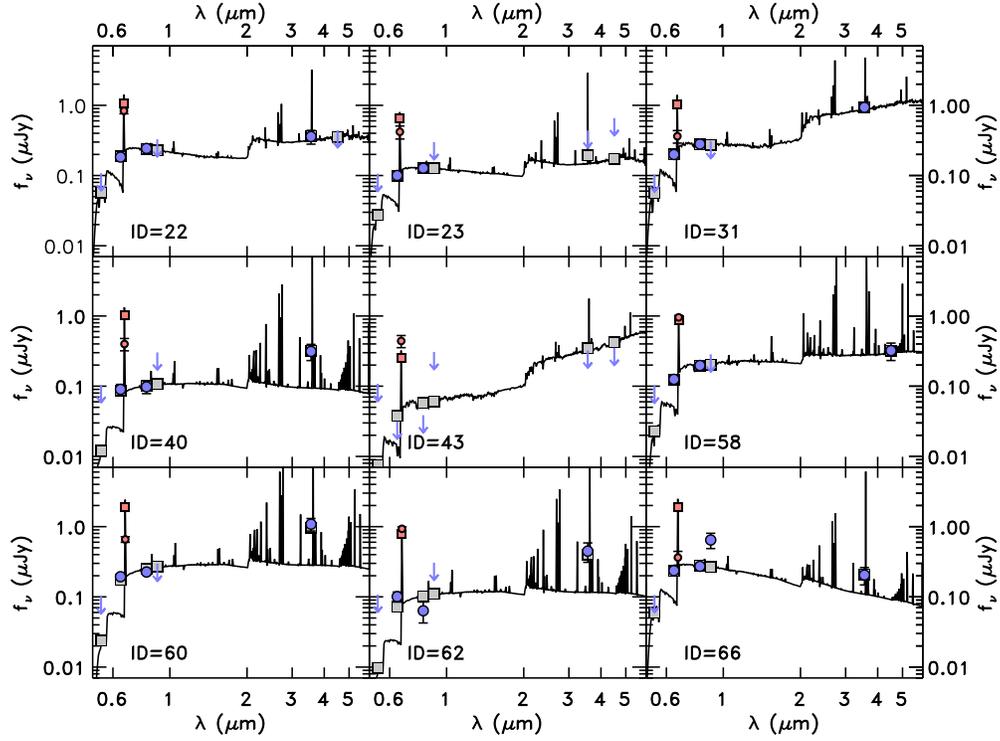}
\vspace{4mm}
\caption{Spectral Energy Distribution fits for individual $z=$ 4.5 LAEs with IRAC and/or HST NIR detections.  The 9 sources shown are from the non-spectroscopically confirmed sample. Observed fluxes are shown by the blue data points, blue arrows are 2$\sigma$ upper limits.  The model average bandpass fluxes are shown by the grey squares.  When performing the SED fitting we allowed for varying star formation histories, varying amounts of dust, metallicity, and nebular emission.}
\end{figure*}

\begin{figure*}
\epsscale{0.8}
\plotone{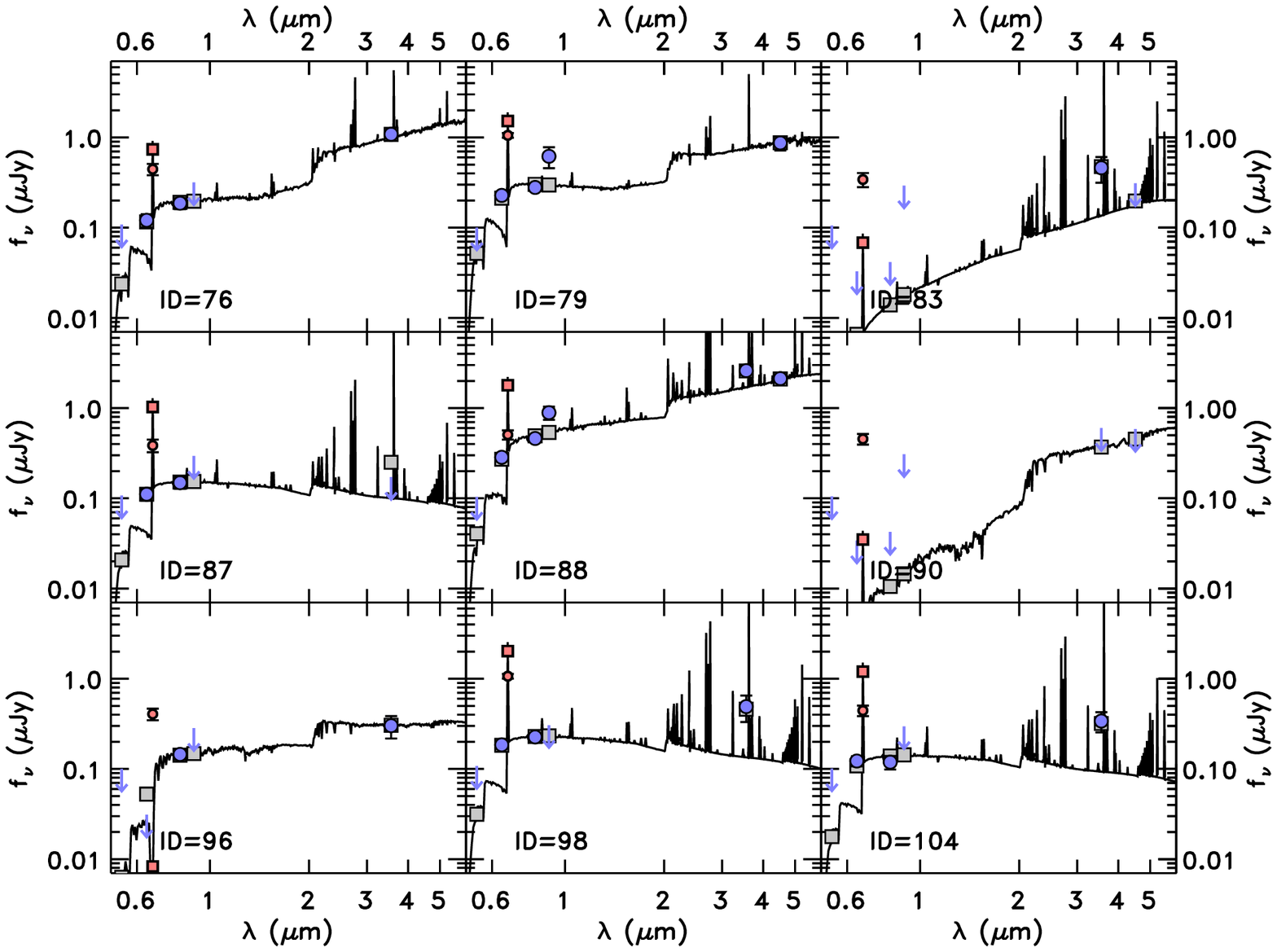}
\caption{Same as Figure 12}
\vspace{8mm}
\end{figure*}

\begin{figure*}
\epsscale{0.8}
\plotone{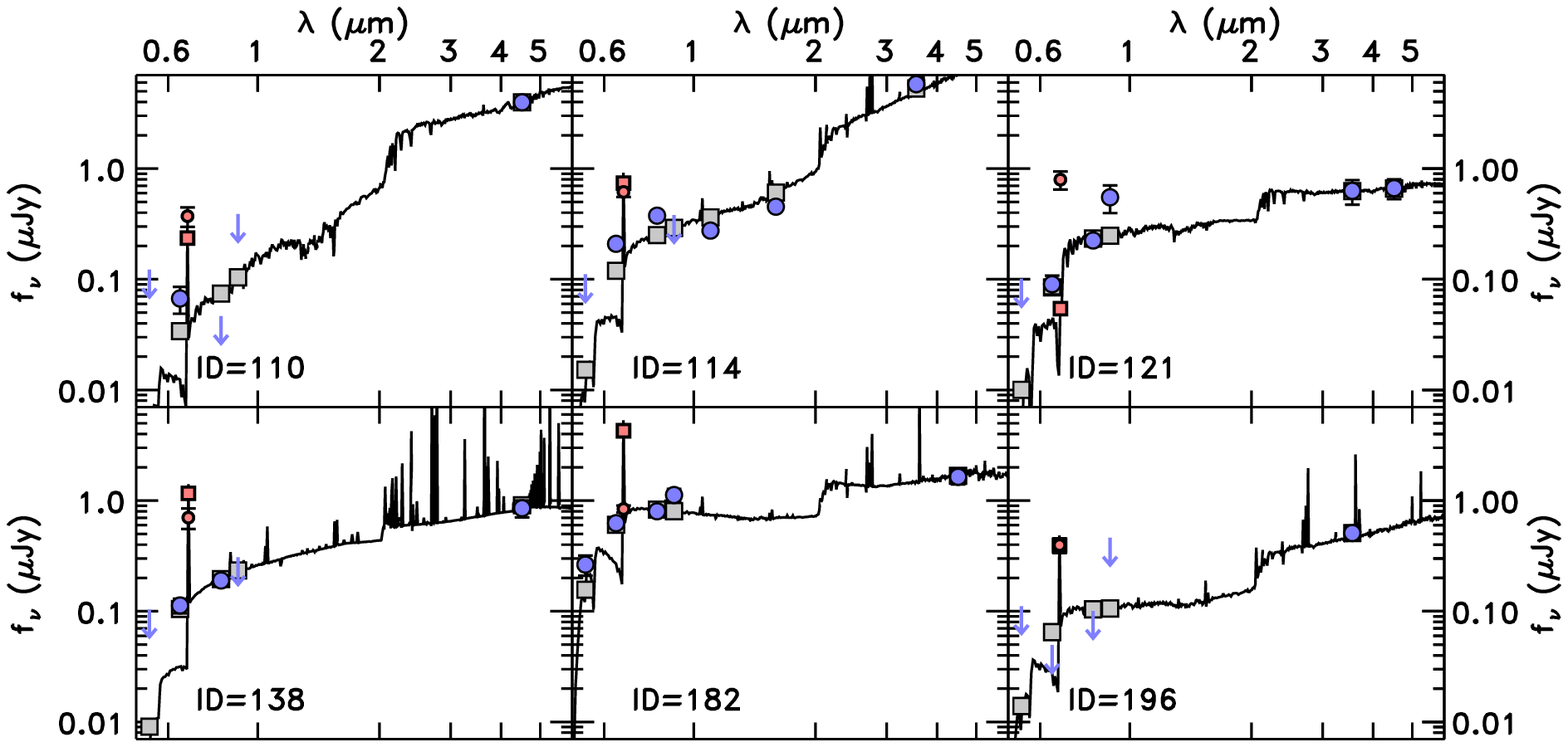}
\vspace{-25mm}
\caption{Same as Figure 12}
\end{figure*}

\begin{figure*}
\epsscale{0.8}
\plotone{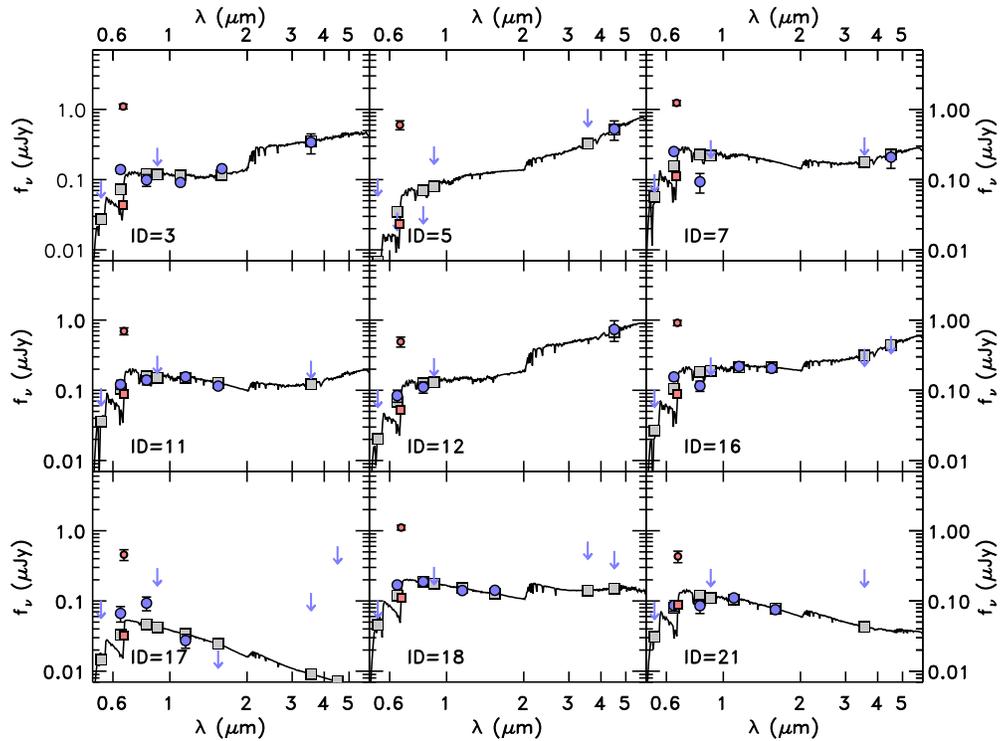}
\vspace{4mm}
\caption{Spectral Energy Distribution fits for individual $z=$ 4.5 LAEs with IRAC and/or HST NIR detections.  The same sources as are shown in Figure 9, this time without allowing for nebular emission lines in the fitting.}
\vspace{4mm}
\end{figure*}

\begin{figure*}
\epsscale{0.8}
\plotone{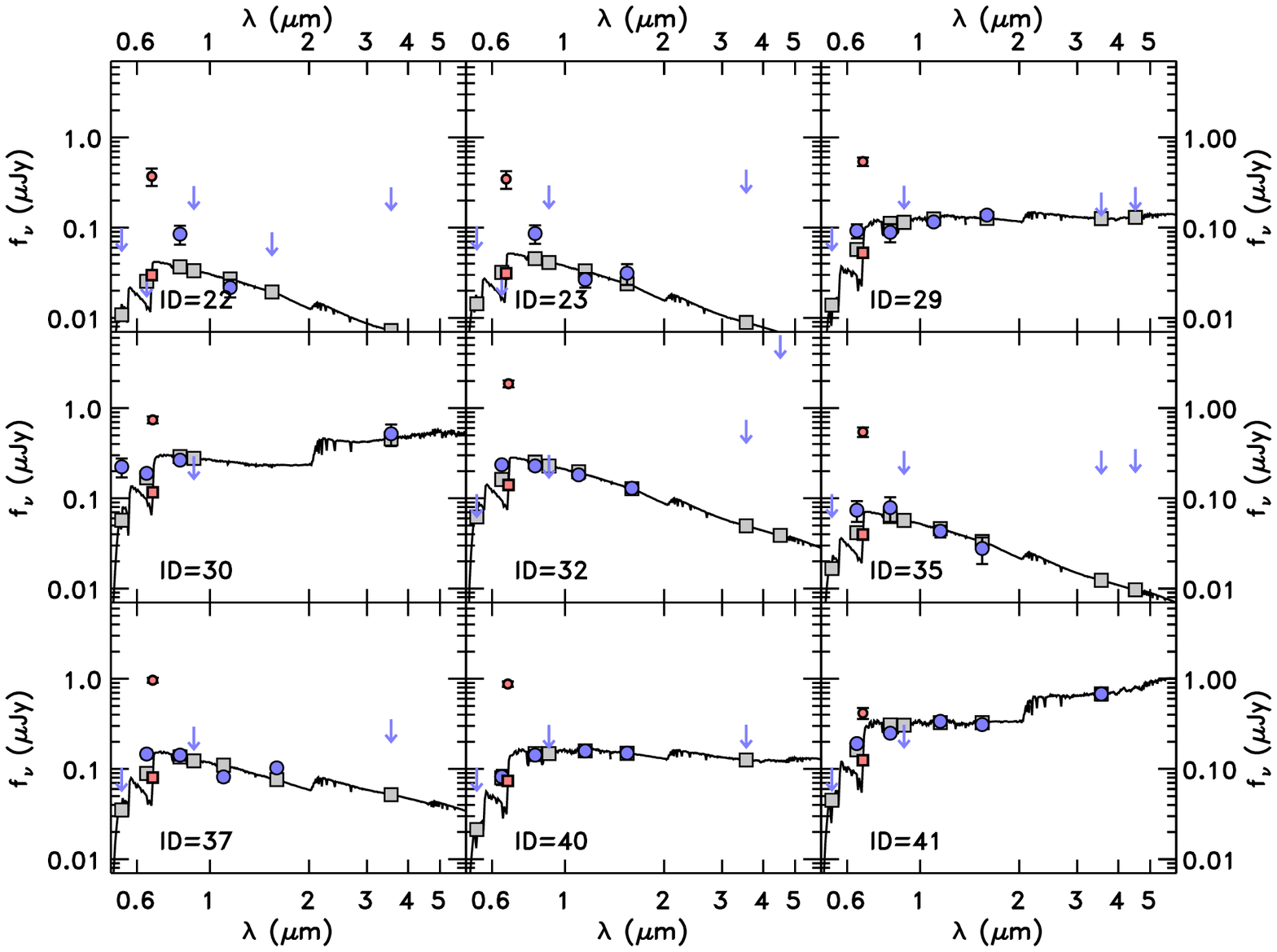}
\vspace{4mm}
\caption{Same as Figure 15; showing same sources as in Figure 10 except without nebular emission.}
\end{figure*}

\begin{figure*}
\epsscale{0.8}
\plotone{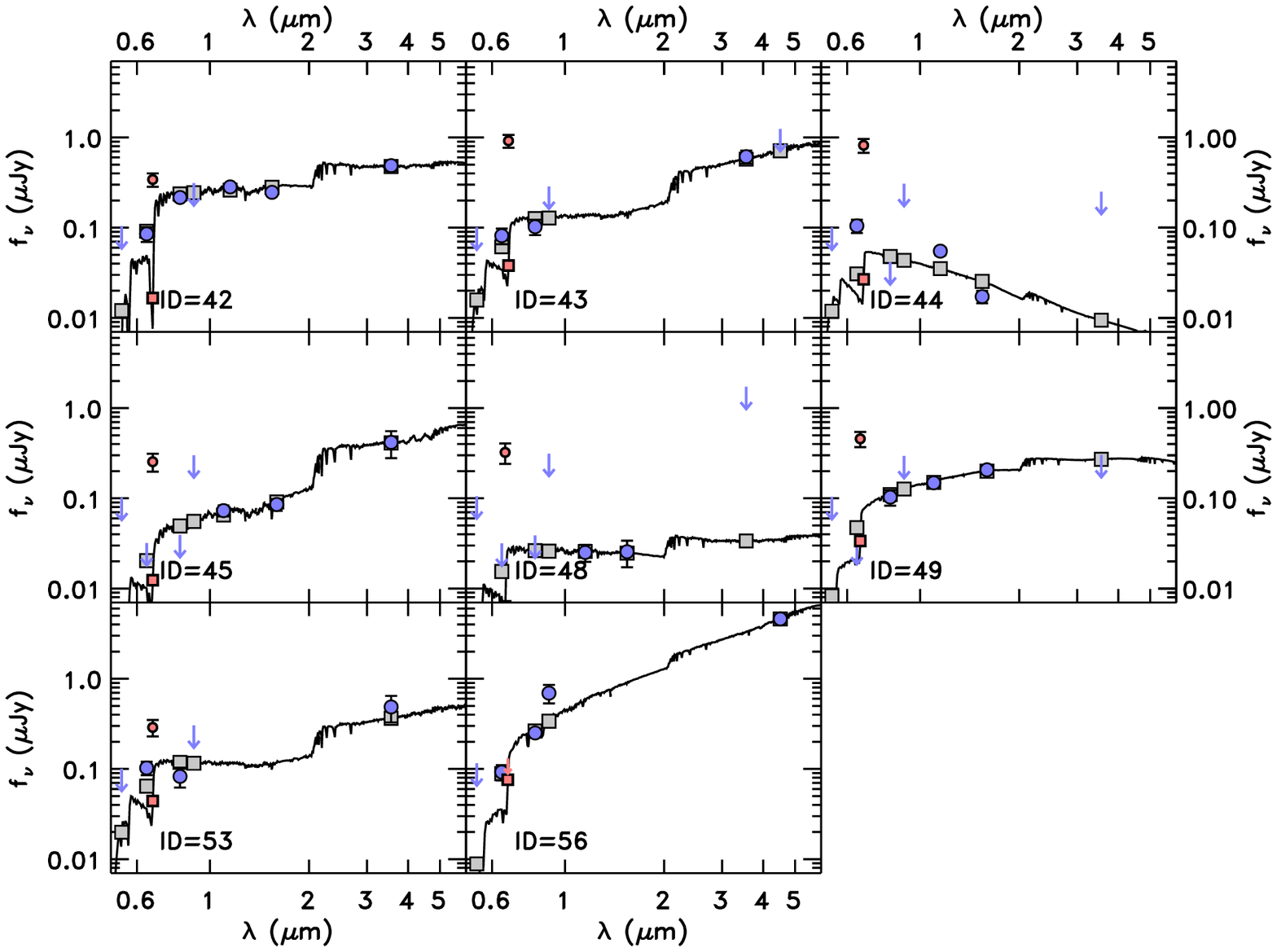}
\vspace{4mm}
\caption{Same as Figure 15; showing same sources as in Figure 11 except without nebular emission.}
\end{figure*}

\begin{figure*}[!t]
\epsscale{1.0}
\plotone{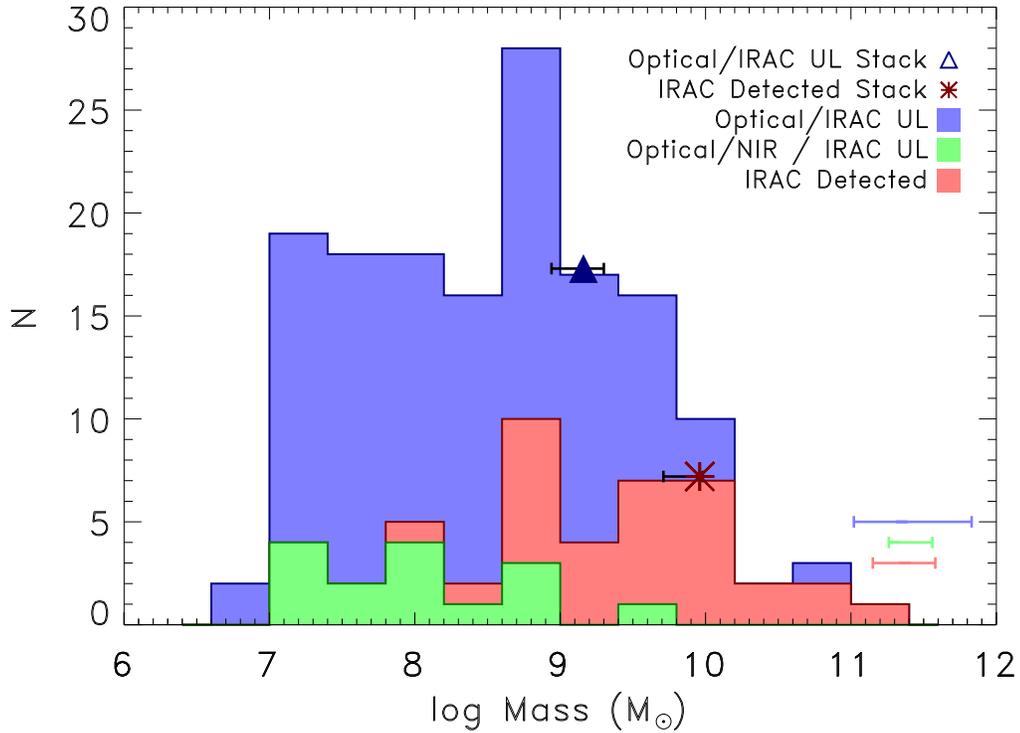}
\caption{Stellar Mass histogram for all candidate $z=$ 4.5 LAEs based on the best-fit SED fitting and 1000 Monte Carlo simulation for each galaxy. The red histogram denotes the mass distribution of LAEs that have at least 3$\sigma$ flux detections in at least one IRAC band, some of these objects are also detected in the NIR.  The green histogram shows the mass distribution of LAEs that have 3$\sigma$ flux detections in at least one NIR band, but only have upper limits in the IRAC bands.  The blue histogram is for objects with detections only in the optical, but with upper flux limits in one or both IRAC bands. The average 68$\%$ confidence ranges for each sample are shown by the error bars based on the Monte Carlo simulations. The IRAC and NIR detected samples have typical mass errors of 0.2 dex, as compared to the non-IRAC detected sample, where the stellar mass uncertainties are closer to 0.4 dex.  The addition of the longer wavelength data does improve the SED fitting results, most explicitly for stellar mass. The best-fit mass and the 68$\%$ confidence range for the two stacks (IRAC-detected; IRAC-undetected) are also over-plotted and are denoted by the asterisk and triangle, they are plotted at the appropriate mass value, but are placed at an arbitrary value on the y-axis, above the corresponding red or blue histograms.}
\end{figure*}

\begin{figure}
\epsscale{1.1}
\plotone{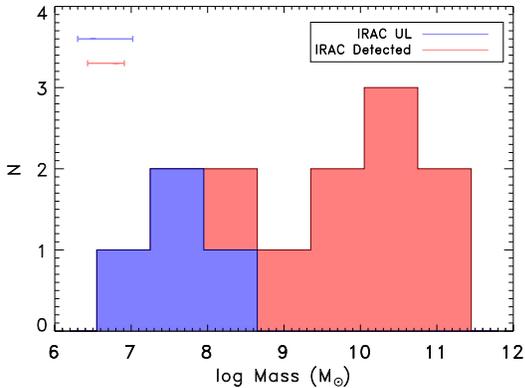}
\caption{Mass histogram for the candidate $z=$ 5.7 LAEs based on the best-fit SED fitting and 1000 Monte Carlo simulation for each galaxy. The red histogram denotes the mass distribution of the nine LAEs that are detected in IRAC.  The blue histogram is for objects with detections only in the ground-based optical, but with upper flux limits in one or both IRAC bands. The average 68$\%$ confidence ranges for each sample are shown by the error bars based on the Monte Carlo simulations.}
\end{figure}

\begin{figure}
\epsscale{1.1}
\plotone{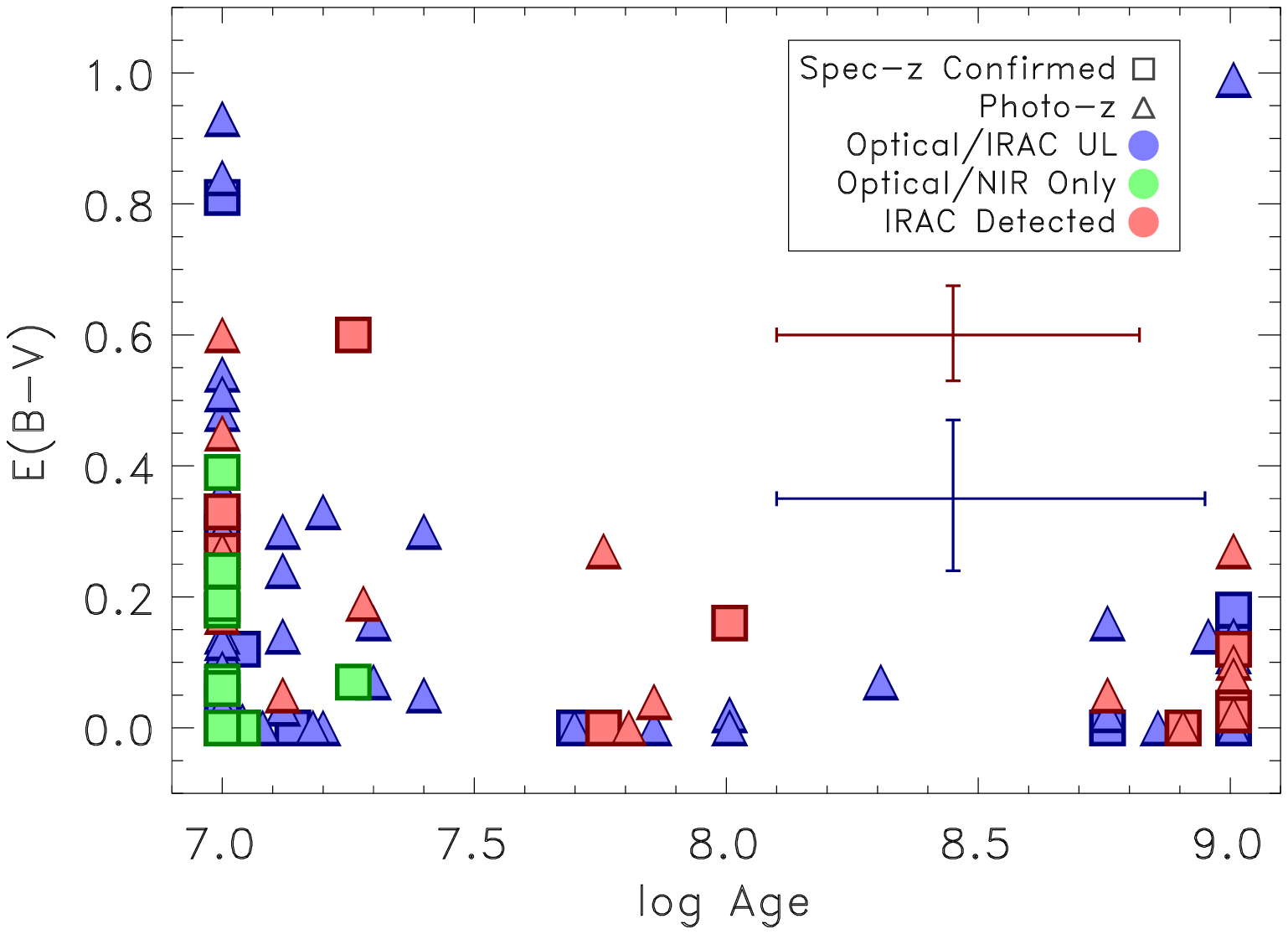}
\caption{Best-fit age and dust estimates for the $z=$ 4.5 LAE sample.  Squares show the spectroscopic confirmed sources, whereas the triangles show sources that have only been selected via their narrowband imaging. Red points denote LAEs detected in IRAC, green points are LAEs detected in the NIR and optical only, although some have upper limits from IRAC, and blue points are LAEs only detected in the optical but with upper limits from IRAC.   The typical errors bars for the different samples (IRAC and/or NIR detected vs. non-IRAC detected) are plotted.  We see a wide spread in both age and dust parameters, but the typical errors for individual galaxies are large, except in a handful of cases, as these two parameters are not as well constrained as the stellar mass estimates from SED fitting.}
\end{figure}

\subsubsection{Stacking Results}
The best-fit SEDs fit to the data for the two stacked samples are shown in Figure 21. In this figure we also show the range for 100 of the best-fit SEDs from the Monte-Carlo simulations (light grey lines).  By stacking the data, we again confirm that the IRAC-detected objects on average are more massive than those that are undetected in the IRAC bands.  We obtain a median best-fit stellar mass of 1.5$\times$10$^{9}$ M\sol~(0.86 -- 2.0$\times$10$^{9}$ M\sol~68\% confidence) for the IRAC-undetected stack, and best-fit mass of 9.2$\times$10$^{9}$ M\sol~(5.1 -- 9.5$\times$10$^{9}$ M\sol~68\% confidence) for the IRAC-detected stack.  This stacking analysis demonstrates that the addition of the median fluxed combined stacked IRAC data helps better constrain the range of models, as seen by the smaller spread in stellar mass for the IRAC-undetected stack as compared to the individual IRAC-undetected LAEs.  The IRAC-detected stack does appear to have a slightly better constrained fit on stellar mass than the IRAC-undetected stack, shown by the smaller spread in best-fit SEDs plotted in the left-hand panel of Figure 20, as well as a somewhat smaller range on the stellar masses.  

The range in other parameters, such as age and dust, do not appear to be much better constrained in the IRAC-detected stack as compared to the IRAC-undetected stack.  The SED fitting results from the stacking analysis are summarized in Table 7.  From the Monte-Carlo simulations, we measure a 68$\%$ confidence range in best-fit values of E(B -- V) of 0.01 -- 0.17 and 0.00 -- 0.09 for the IRAC-detected stack, and IRAC-undetected stack, respectively.  We measure a 68$\%$ confidence range for age of 203 -- 1015 Myr and 64 -- 570 Myr for the IRAC-detected stack, and IRAC-undetected stack, respectively.  Stacking thus allows us to place some constraints on the age and dust attenuation, even in low-mass galaxies which were individually undetected with IRAC.  However, we find here (similar to previous works, e.g. Nilsson et al.\ 2011; Pirzkal et al.\ 2012; Aquaviva et al.\ 2013) that even when stacking, we still obtain tighter constraints on the stellar mass than other physical properties.

\begin{figure*}[!t]
\epsscale{1.1}
\plottwo{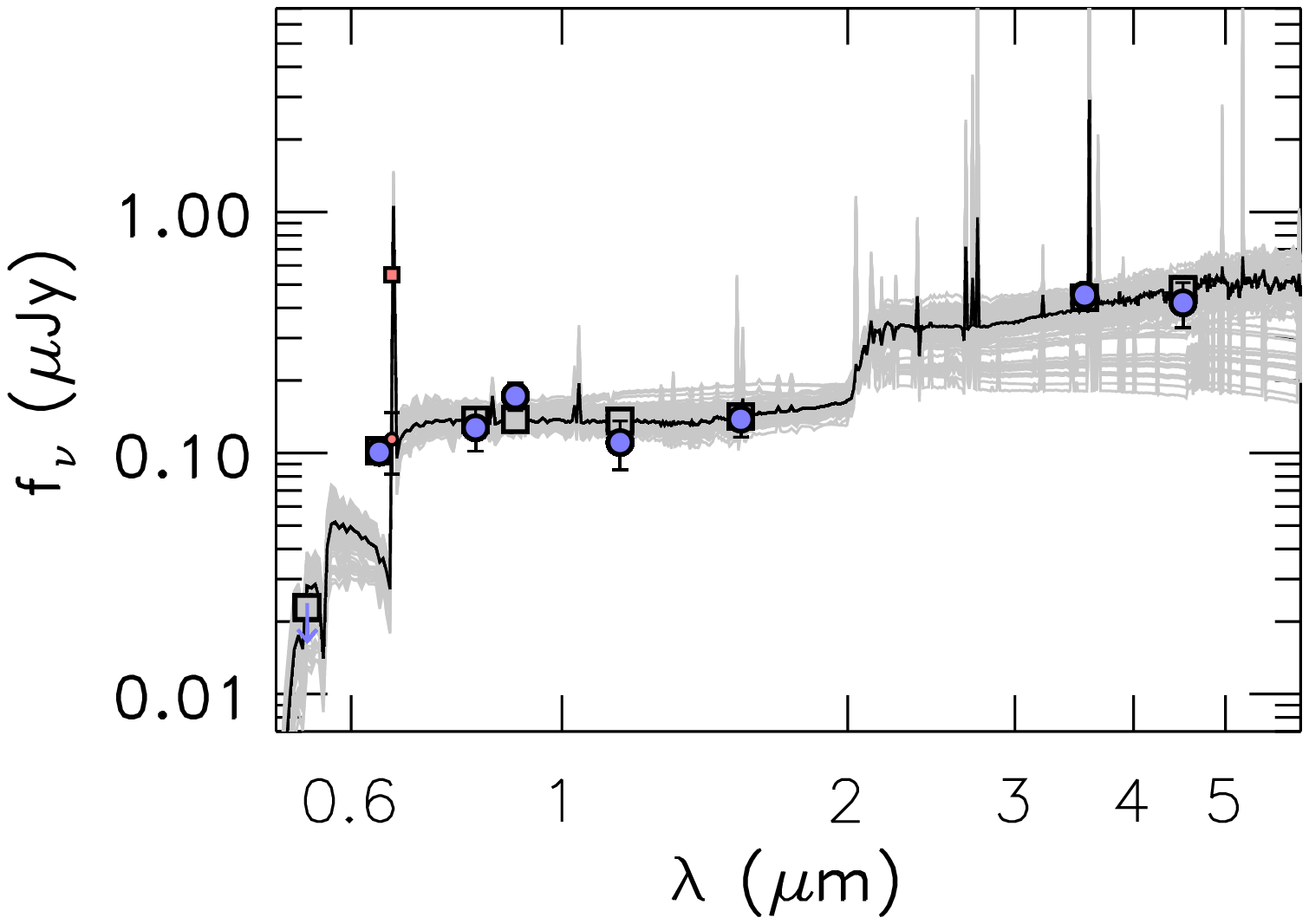}{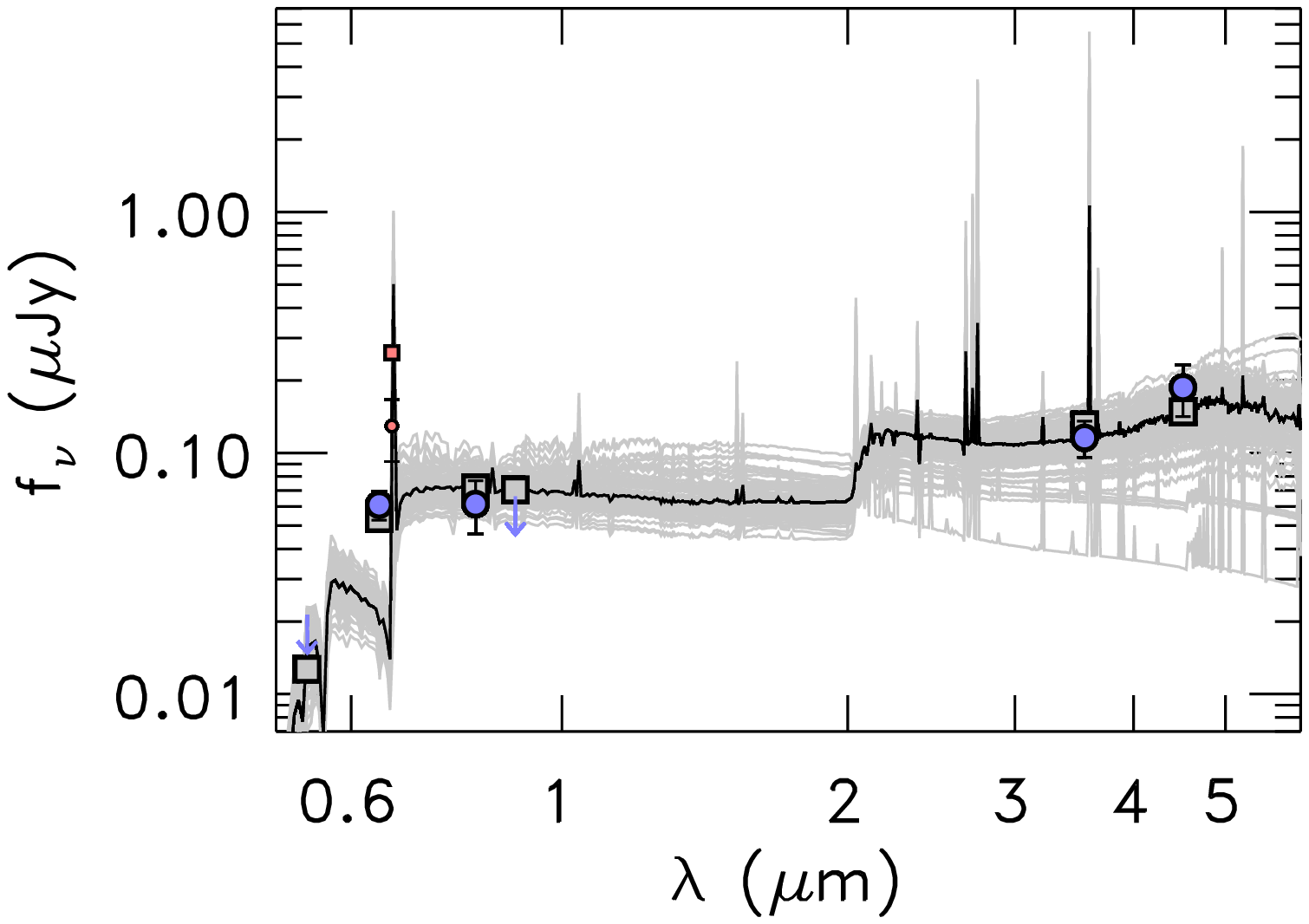}
\caption{SED fits for the $z=$ 4.5 LAEs stacked fluxes. Left: Stacked SED for the IRAC detected sample.  Right: Stacked SED for the undetected IRAC sample. Observed fluxes are shown by the blue data points, blue arrows are 2$\sigma$ upper limits.  The model average bandpass fluxes are shown by the grey squares.  The light grey lines show the best-fits from 100 Monte-Carlo simulations. The addition of the IRAC detections helps better constrain the range of models, most specifically the stellar mass, as compared to the IRAC-undetected stack.  When performing the SED fitting we allowed for varying star formation histories, varying amounts of dust, and nebular emission.}
\end{figure*}

\section{Discussion}
\subsection{Stellar Masses -- Comparison to other LAE studies}
Other recent studies of LAEs and LBGs at similar redshift have used {\it Spitzer} IRAC observations to try to better constrain the stellar masses and other properties of z $\sim$ 4 galaxies.  We compare our derived $z=$ 4.5 LAEs stellar population properties to previous studies of LAEs that have made use of IRAC observations.  The Lai et al.\ $z=$ 3.1 sample was stacked into an IRAC-detected and IRAC-undetected sample, with the IRAC detected sample having an average mass of 9$\times$10$^{9}$ M\sol, and the undetected sample having an average mass of 3$\times$10$^{8}$ M\sol.  This is comparable to our $z=$ 4.5 results, with the IRAC-detected stack being more massive at both redshifts, with our $z=$ 4.5 IRAC-detected stack having a best-fit stellar mass of 9$\times$10$^{9}$ M\sol, and our IRAC-undetected stack having a best-fit stellar mass of 15$\times$10$^{8}$ M\sol.  Individually, when we compare our IRAC-detected objects to those fit by Finkelstein et al.\ (2008) at the same redshift, we find a similar overlap for some of our IRAC-detected sample, but with a fraction of our sample having higher masses.  Of those detected in at least one IRAC band in the $z=$ 4.5 LAE sample from Finkelstein et al.\ (2009), they find a range of stellar masses of 3$\times$10$^{8}$ -- 6$\times$10$^{9}$ M\sol, we find LAEs in this same mass range in our sample, however we also have 11 galaxies with best-fit masses greater than 10$^{10}$ M\sol, representing 7$\%$ of our total sample.  Of our massive LAEs, 20$\%$ come from the spectroscopically confirmed sample; this is somewhat similar to the overall sample breakdown between spec-z galaxies and non-spec-z galaxies, with 30\% of the total galaxies observed with IRAC being a spec-z confirmed LAE.  Detecting a higher fraction of massive LAEs as compared to Finkelstein et al.\ (2009) is likely due to the much larger survey area of our study, allowing for detection of a fraction of LAEs that represent the rare, most massive, more evolved population of galaxies at this redshift.  However, it is possible that some of these massive galaxies which are not yet spectroscopically confirmed could be lower redshift interlopers, though the fraction is likely small, given the overall $\sim$75\% spectroscopic confirmation success rate (Dawson et al.\ 2007).

\subsection{H$\alpha$ Emission}
In addition to better constraints on stellar mass, the {\it Spitzer} IRAC data can tell us about possible contribution to the SED from H$\alpha$ emission.  There are clearly cases in our sample where there is a high 3.6$\mu$m band-flux, and the best-fit template fits this with a strong H$\alpha$ flux.  Examples of this include sources 43c, 53c (Fig.\ 11); and 40nc, 60nc, 62nc (Fig.\ 12).  In these sources, the observed narrowband photometry is also a good match to the predicted Ly$\alpha$ flux from the best-fit templates, even though the narrowband fluxes were not used in the fits.  This implies that despite the uncertainties in the radiative transfer of Ly$\alpha$, our simple assumptions regarding Ly$\alpha$ escape discussed in \S3.1 may be appropriate. 

In a recent study, Stark et al.\ (2013) derive an H$\alpha$ equivalent width distribution for their sample of 92 spectroscopically confirmed LBG galaxies at $z=$ 3.8 -- 5, based on the 3.6 $\mu$m flux.  They find an average rest-frame H$\alpha$ equivalent width of 270 \ang, indicating that nebular emission contributes at least 30$\%$ to the 3.6 $\mu$m flux.  Shim et al.\ (2011) also have a sample of 74 z$\sim$4 LBGs detected in {\it Spitzer} IRAC 3.6 and 4.5 $\mu$m observations.  They show that 70$\%$ of their sources show an excess in 3.6$\mu$m over the stellar continuum.  In this sample of LAEs we have 22 sources with IRAC 3.6 $\mu$m detections above 3$\sigma$, of those 22 sources 7 are from the spectroscopically confirmed sample, and the other 15 are from the narrowband-selected sample.  Of these 22 sources, 11 have 3.6 $\mu$m emission above the stellar continuum as determined from the best-fit SED.  This corresponds to 50$\%$ of this subset sample having H$\alpha$ emission above the stellar continuum, somewhat comparable to the 70$\%$ in the Shim et al.\ sample.  However, the presence of an emission line (i.e. a 3.6 $\mu$m  excess) implicitly makes a given object easier to detect with IRAC.  A conservative lower limit to the fraction with strong H$\alpha$ emission would be $>$9\% (11 with a 3.6 $\mu$m excess compared to 123 total 3.6 $\mu$m-observed galaxies).

We also see when only looking separately at the spectroscopically confirmed sample, and the non-spectroscopically confirmed sample, that both samples of IRAC-detected objects have similar fractions of 50$\%$ with 3.6 $\mu$m detections have emission above the stellar continuum.  This similar fraction of an excess in  3.6$\mu$m over the stellar continuum in both our samples gives credence to our narrowband only selected LAEs, indicating both samples are probing similar properties of LAEs at this redshift.  While a smaller sample fraction, this subsample of spectroscopically confirmed LAEs may be a robust estimate of the true LAE population as their redshift is known, and a higher fraction of them also have HST NIR detections as well, resulting in better estimates on the SED fitting (in particular, on the continuum blue-ward of H$\alpha$).  Given the strong Ly$\alpha$ emission from our sample of LAEs, we would expect strong H$\alpha$ as well.

\subsection{Star Formation Rates vs.\ Stellar Mass}
Previous studies at lower redshift have shown that there is a sequence of star forming galaxies, with a nearly linear relationship between star formation and stellar mass, known as the ``main sequence (MS) of star formation'' (e.g. Noeske et al.\ 2007; Daddi et al.\ 2007; Elbaz et al.\ 2007) at a given redshift. These studies have shown that the tight relations exist locally and at redshifts 1 and 2, and that the slope of the trend does not seem to evolve much with redshift.  The sequence normalization does vary, with lower normalizations at lower redshift, demonstrating that higher--redshift star forming galaxies are forming stars at higher rates compared to similar mass galaxies at lower redshift.  More recent work has shown that these trends continue to higher redshift. Weinzirl et al.\ 2011 showed that a sample of more massive galaxies at redshift 2 -- 3 continues to follow the Daddi $z=$ 2 trend.  Hathi et al.\ 2013 looked at a sample of LBGs at $z=$ 1 -- 3 and a comparison sample at $z=$ 4 -- 5, with both samples showing a higher normalization above the Daddi $z=$ 2 MS, though both samples are best-fit by a trend line with a logarithmic slope of 0.9, similar to z $\sim$ 2 samples from both Daddi et al.\ (2007) and Sawicki (2012).  Even more recently, Speagle et al.\ (2014) have investigated the MS out to $z \sim$ 6 and find that the width of the MS distribution remains constant over time, with a spread of $\sim$0.2 dex.  They also note that the scatter around the MS at a fixed mass may be due to scatter in time, i.e., an uncertainty in the age of the universe at a given mass and star formation rate.

We investigate our sample of $z=$4.5 LAEs by placing them on the star formation rate (SFR) versus stellar mass plane.  We calculated the SFR assuming the Kennicutt (1998) UV star formation calibration defined from the dust-corrected UV absolute magnitude from 1500 -- 2800 \ang, assuming a Salpeter IMF (as in Kennicutt 1998).  Using the best-fit model for each galaxy, we calculated the model UV flux from 1500 -- 2800 \ang, and then corrected for dust attenuation by taking the best-fit E(B-V) dust estimate and converting it to a dust extinction at A$_{2150}$ (assuming a Calzetti dust law).  The 68$\%$ confidence range on the dust parameters was used in determining uncertainties on the derived SFRs.  The best-fit mass and derived SFR are plotted in Figure 22, with error bars based on the Monte Carlo simulations for each quantity.  

We find that our sample of z$\sim$4.5 LAEs follow a similar linear correlation, though the normalization is higher than the z$\sim$2 derived MS of star forming galaxies from Daddi et al.\ (2007), who also assume a Salpeter IMF.  Interestingly, we do find that the majority of our massive (log M/M\sol\ $>$ 9.5) LAEs fall directly on the $z \sim$ 2 trend.  Fitting to the IRAC-detected and HST-detected LAEs between the mass range of 10$^{7}$ -- 3$\times$10$^{9}$, we derive a best-fit line with a logarithmic slope of 0.9 $\pm$ 0.05.  In Figure 22, the solid cyan line shows this best-fit trend for our sample, extrapolated out to higher masses.  The dashed cyan lines show the scatter from the best-fit line, which is $\sim$0.3 dex for the LAEs at $z=$ 4.5.  This is similar to the spread measured by Daddi et al.\ of $\sim$0.3 dex at $z=$ 2, and by Hathi et al.\ of $\sim$0.4 dex for LBGs at $z=$ 4 -- 5, although higher than the overall spread measured by Speagle et al.\ across multiple epochs.  

For comparison, we plot the Daddi et al.\ $z =$ 2 (blue line), the Hathi et al.\ (green line) and the Speagle et al.\ (red line) trends at $z =$ 4 MS relations alongside our data in Figure 22.  At similar redshift, our best-fit is elevated above the trend found by Hathi et al.\ for continuum selected galaxies, but is consistent within 2$\sigma$.  Our results are also somewhat consistent with those of Bouwens et al.\ (2012), who found an average Mass--SFR relation for a sample of $z=$ 4 dropout galaxies with a logarithmic slope of 0.73 $\pm$ 0.32, and with a normalization similar to that of the Hathi et al.  However, our sequence is elevated even more above the $z =$ 4 trend determined by Speagle et al.\ (2014) by a factor of 4--5.   A recent study by Schreiber et al.\ (2014) using Herschel observations and stacking a sample of massive galaxies at $z=$ 3.5--5, with an average stellar mass of 2$\times$10$^{11}$ M\sol, finds an average SFR of $\sim$630 M\sol~yr$^{-1}$ for that stack. This sample is at the upper mass end of our sample of LAEs, and appears to fall in between our two LAE populations on the SFR -- stellar mass plot (orange square in Figure 21), but is consistent with the Hathi et al.\ LBG sample at these redshifts.  However, another recent study of the SFR -- stellar mass relation by Salmon et al.\ (2015), looking at galaxy samples at redshifts 4 -- 6, found for their $z=$ 4 sample, a logarithmic slope of 0.7 $\pm$ 0.21, but with a lower normalization than what we see for our main $z=$4.5 LAE sample.  The Salmon et al.\ sample has a normalization more consistent with the Daddi et al.\ results. Our $z=$ 4.5 LAE sample used to estimate the Mass--SFR relation is $\sim$3.5$\sigma$ from the Daddi et al.\ $z=$ 2 MS.  Overall, we find that for LAEs with stellar masses $<$3$\times$ 10$^{9}$ M\sol, the SFR-stellar mass relation of LAEs at $z =$ 4.5 is similar in slope and spread but is elevated in normalization by 4--5 times compared to continuum-selected galaxies at the same redshift.  Other recent studies of LAEs at redshifts 2 -- 3 have also found that the general LAE population lies above the normal z$\sim$2 star forming galaxy main sequence (Song et al.\ 2014; Vargas et al.\ 2014).

\begin{figure*}[!t]
\epsscale{1.0}
\plotone{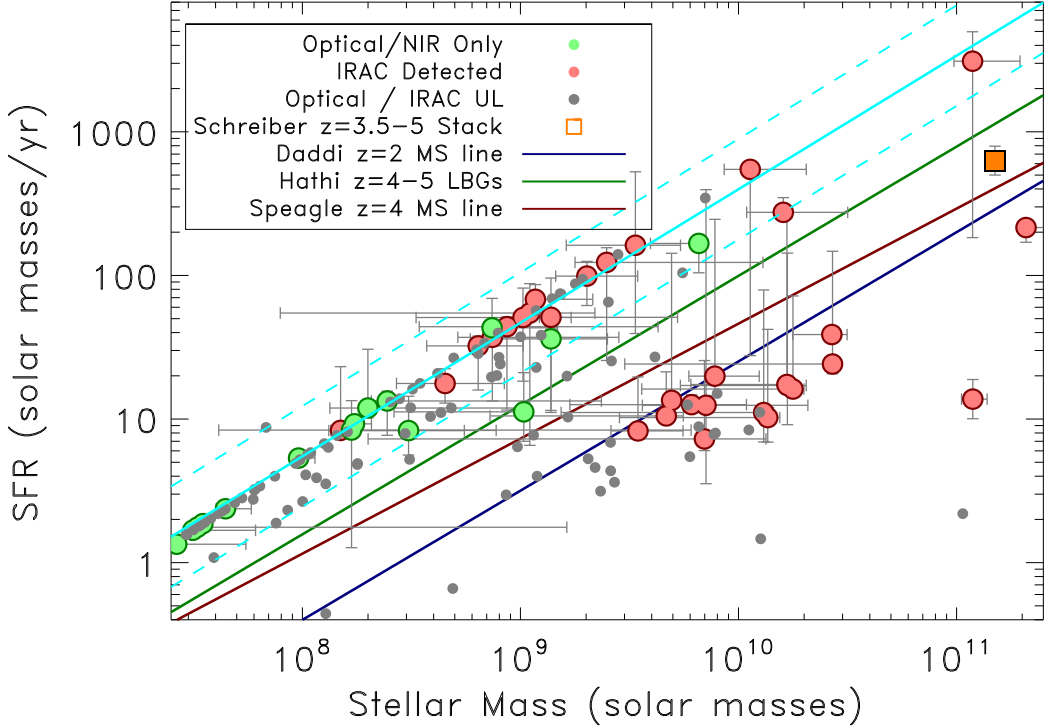}
\caption{Best-fit stellar mass versus the dust-corrected UV SFR, derived assuming the Kennicutt UV SFR calibration from 1500--2800 \ang.  Error bars on each quantity are based on the 1000 Monte Carlo simulations for stellar mass and dust.  The observed z$\sim$2 trend for star forming Main Sequence galaxies from Daddi et al.\ (2007) is shown by the blue line, while the green line is observed trend for $z=$ 4--5 LBGs from Hathi et al.\ (2013), and the red line is the $z=$ 4 MS trend from Speagle et al.\ (2014). The red points are the observed LAEs with IRAC detections (3$\sigma$), the green points are for LAEs with NIR detections, and the small grey points are for the LAEs with only optical detections + IRAC upper limits. For comparison, a stack of galaxies from $z=$ 3.5--5 with Herschel observations from Schreiber et al.\ (2014) is denoted  by the orange square. For the most part, the LAEs lie above the z$\sim$2 trend line, but follow a similar shape and trend.  A smaller fraction of the LAE population at z$\sim$4.5 does fall right on the Daddi et al.\ line. The solid cyan line is the best-fit line to the IRAC-detected and HST-detected LAEs between the mass range of 10$^{7}$ -- 3$\times$10$^{9}$, extrapolated out to higher mass. This best-fit line has a logarithmic slope of 0.90 for our sample.  The dashed cyan lines show the scatter from the best-fit line, which is $\sim$0.3 dex for the LAEs at $z=$ 4.5.}
\end{figure*}

To better investigate the two different populations of $z=$ 4.5 LAEs seen on the SFR versus stellar mass plot, we looked at individual stellar properties of IRAC-detected objects that have similar masses, but substantially different star formation rates.  One set contains objects with SFRs above 100 M\sol~ yr$^{-1}$ and masses between a few $\times$ 10$^{9}$ -- few $\times$ 10$^{10}$.  There are 6 IRAC-detected objects in this set.  The second set contained objects with SFRs less than 50 M\sol~ yr$^{-1}$, but within the same mass range as the first set.  There are 13 IRAC-detected objects in this set.  We find that the lower SFR sample (those that more closely follow the $z=$ 2 MS line) is typically older than the higher SFR sample.  The lower SFR sample has best-fit ages ranging from 60 Myr -- 1 Gyr, with only four galaxies having ages between 60 -- 100 Myr, and all others having ages between 570 -- 1000 Myr; whereas the higher SFR sample has best-fit ages ranging from 10 -- 60 Myr, with only one galaxy in this higher SFR sample having a best-fit age above 20 Myr.  One possible explanation for this, is that we are again seeing two populations of LAEs (similar to Finkelstein et al.\ 2009); one subset of galaxies is relatively young with high amounts of star formation, and the other population is older with less star-formation.  These two populations could represent different paths for Ly$\alpha$ photon escape.  In the younger population, the galaxies are undergoing a burst of star-formation, likely driving strong outflows, which may allow the Ly$\alpha$ photons to shift out of resonance via multiple scatterings in an outflowing ISM (e.g., Verhamme et al.\ 2008).  For the older population, these galaxies are more evolved with lower SFRs, but they have had time to create holes in the ISM, which could allow Ly$\alpha$ photons to directly escape.  

These two populations represent a bimodality in the specific SFR (sSFR) in our sample of LAEs.  Out of the total sample of 150 LAEs which have IRAC constraints, we find that $\sim$75\% have high sSFRs of 1 $\lesssim$ log (sSFR) $\lesssim$ 2, while 24\% have somewhat lower values of $-$0.3 $\lesssim$ log(sSFR) $\lesssim$ 0.6.  These latter 25\% are the lower SFR, older subset of LAEs, which lie on the $z =$ 2 MS on Figure 22.   However, this bimodality is seen in our best-fit values, thus to see whether it is an artifact of our SED fitting, we analyzed the full sSFR distribution using the 1000 Monte Carlo fits for each galaxy.  We find that for a given galaxy in the upper or lower sequence (high or low sSFR, calculated from the best fit), the majority of the Monte Carlo simulations fall into that same sequence. This is especially true for galaxies in the upper sequence: 85$\%$ of all 1000 Monte Carlo simulations for all galaxies in the upper sequence remain in the upper sequence. However, for the lower star-forming sequence of galaxies, typically only in $\sim$60$\%$ of the simulations were they measured to have the same lower range in sSFR, while the remaining 40$\%$ of their simulations resulted in higher calculated sSFRs.  This is shown in Figure 23, where we plot the full range of measured sSFRs using the 1000 MC simulations for each source. The left-hand plot (blue histogram) shows the combined 1000 MC simulations for the 113 sources falling in the upper star forming sequence (log sSFR $\gtrsim$ 1), and the right-hand plot is the same for all 37 sources falling in the lower star forming sequence (log sSFR $\lesssim$ 0.6).  

Our simulations show that for an object that has a best-fit value of the sSFR that places it in the upper sequence, it is very likely to truly have a high sSFR.  However, for galaxies best-fit in the lower sequence, there is a non-negligible chance ($\sim$40\%) that they truly have high sSFRs, consistent with the upper sequence.  This may imply that there is really only one SFR--stellar mass sequence for LAEs at $z =$ 4.5, and that our methods of measuring physical properties scatter some fraction of sources into a lower sequence.  However, even if $\sim$40\% of the sources in the low sSFR bin truly had high sSFRs, that would still leave 22 of our sample of 150 LAEs (14\%) on the lower sequence (consistent with the $\sim$15\% of LAEs from Finkelstein et al.\ 2009).  Therefore it appears at least possible that there are two sequences of LAEs, though we are limited by the relatively small number of sources in the lower star forming sequence. In addition, as mentioned in \S5.1 some of these massive galaxies have not yet been spectroscopically confirmed, and thus could be lower redshift interlopers, lowering the fraction of sources in the lower star-forming sequence, although the relatively high confirmation fraction for LAEs in general implies that a significant contamination fraction is unlikely. Further studies with larger samples may help to better address this issue of whether or not there are two populations of LAEs.

\begin{figure*}[!t]
\epsscale{1.1}
\plottwo{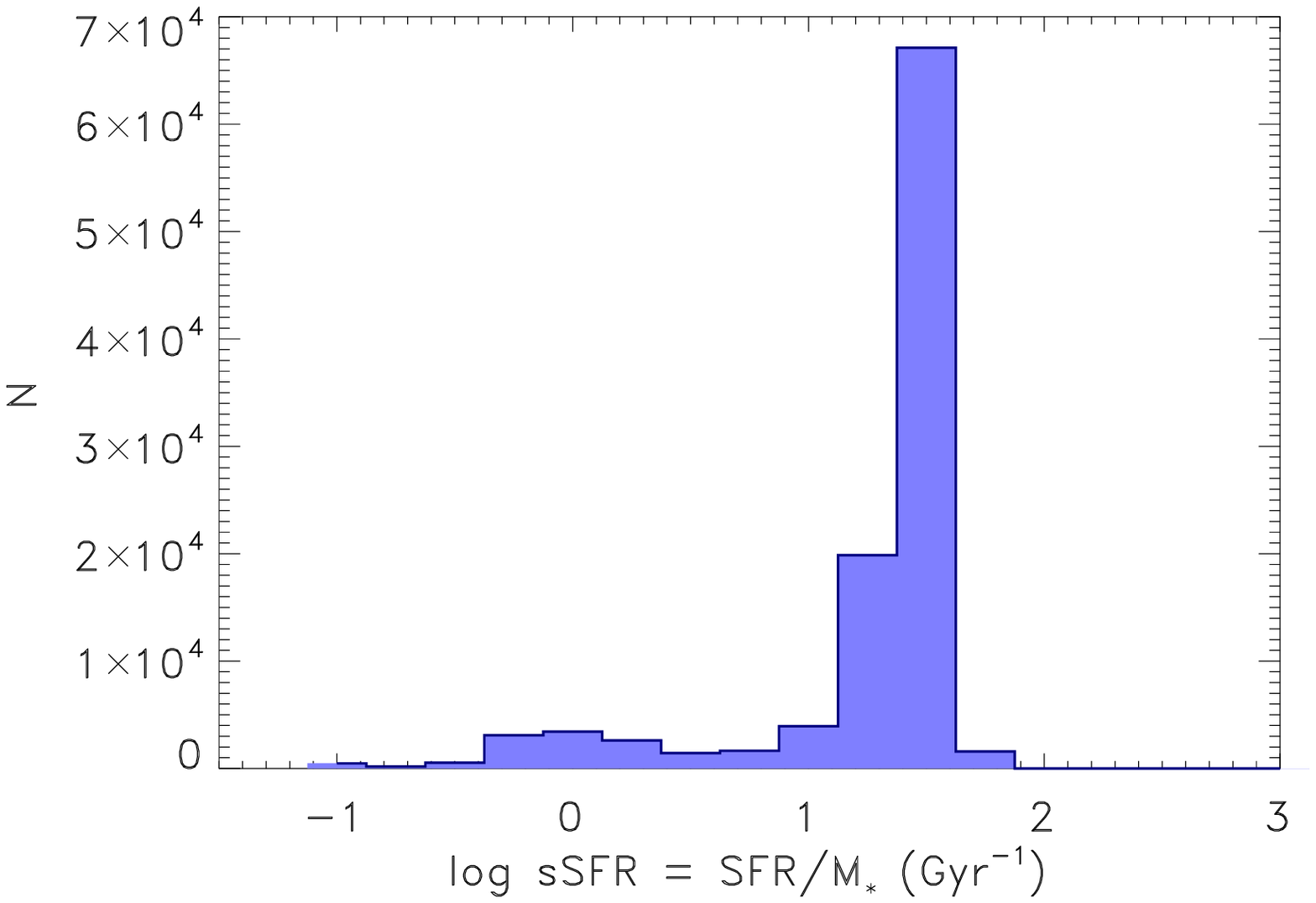}{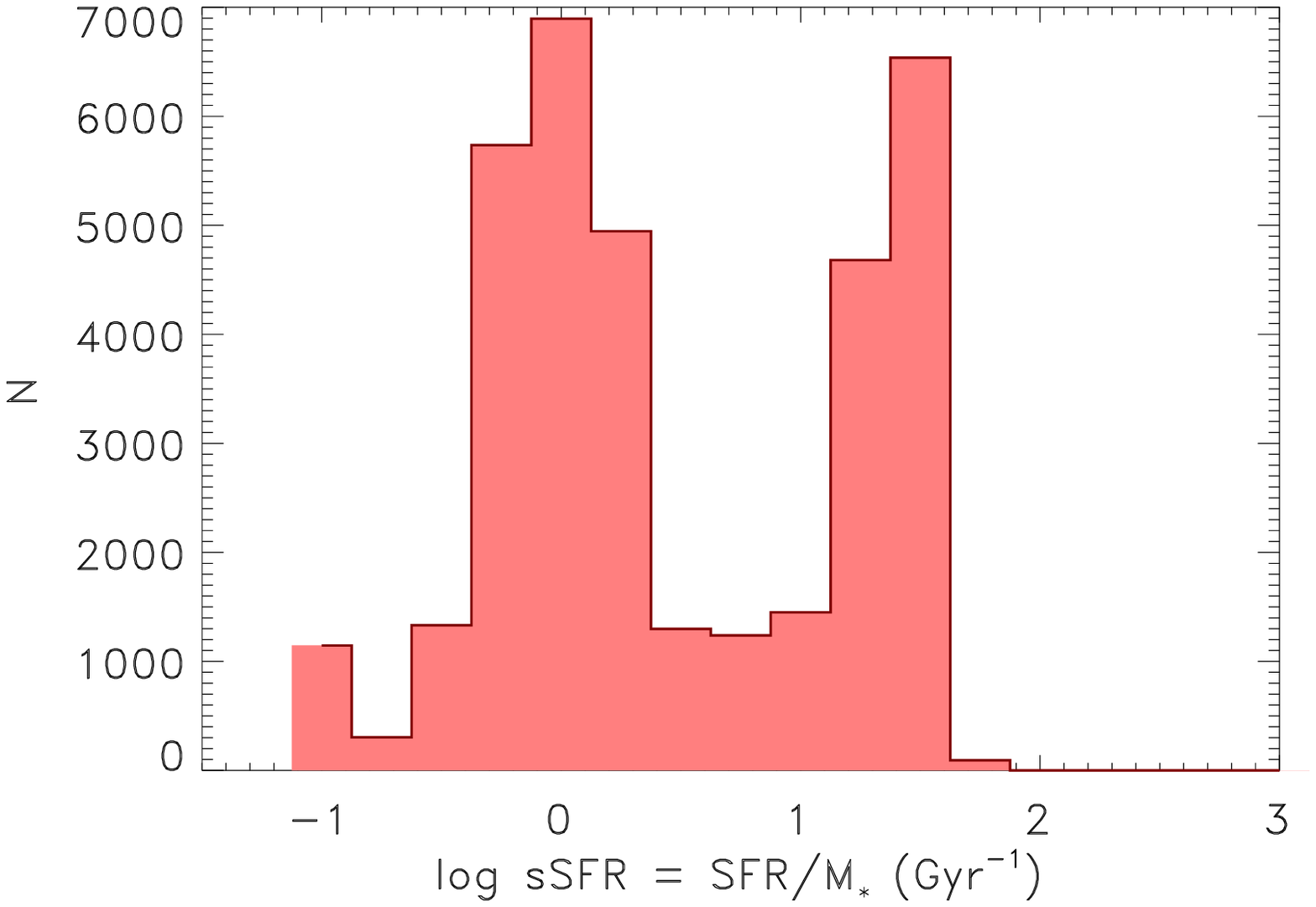}
\caption{The combined 1000 Monte Carlo simulations of specific star formation rate (sSFR) for all sources shown in the star formation -- stellar mass plot from the previous figure. Left: blue histogram is those in the upper star forming sequence (113 sources total). Right: red histogram is for all 37 sources falling in the lower star forming sequence. The left plot shows that sources that fall in the upper sequence are highly confident to be high sSFR sources, and will stay in the upper sequence.  Whereas in the right plot, sources are only likely to remain in the lower star forming sequence 60$\%$, and have a 40$\%$ chance of actually being in the upper sequence.}
\end{figure*}

\section{Summary}
We have investigated a sample of 162 $z=$4.5 LAEs which were observed in the {\it Spitzer} IRAC 3.6 and 4.5 $\mu$m bands. These new {\it Spitzer} observations have allowed for better constraints on the SED fitting of these galaxies, specifically for those detected with IRAC, resulting in better estimates on the stellar masses. 19$\%$ of the $z=$4.5 sample were detected in at least one IRAC band, and this population of LAEs is typically more massive than the LAEs that were not detected in IRAC.   When we fit the IRAC-detected galaxies individually we find typical stellar masses range 5$\times10^{8}$ -- 10$^{11}$ M\sol.  The stellar ages and dust attenuation are more difficult to constrain, even with individual SED fitting and the addition of IRAC data.  Only a few individual objects, typically those with additional photometry data from {\it HST} NIR , have more robust contraints on the stellar age. In general the IRAC detected objects show a wide range in best-fit values for stellar age, ranging from a 10 Myr to 1 Gyr.  By analyzing the physical properties of the individual galaxies that have IRAC detections, as well as performing stacking analyses on the IRAC-detected and IRAC-undetected samples, we find the IRAC-undetected sample has an average stellar mass of $\sim$1.5$\times$10$^{9}$ M\sol, and an average age of 200 Myr.  Whereas, the stacked IRAC-detected sample has an average stellar mass of 9$\times10^{9}$ M\sol, and an average age of 900 Myr.  The stellar mass and age estimates found for this sample agree with previous studies of LAEs at this redshift, however we are also detecting a population of relatively massive LAEs at this redshift, with a handful of our IRAC-detected sample having masses between a few $\times$ 10$^{10}$ -- 10$^{11}$ M\sol. 
We have also used and demonstrated the importance of including nebular emission in the stellar population fitting process.  When comparing best-fit ages of individual LAEs fit with and without nebular emission lines, we find a mean age ratio of  Age$_{neb}$ / Age$_{no ~neb}$ $=$ 0.5.  This indicates that an individual LAE is 50$\%$ more likely to have a best-fit stellar population age greater than 600 Myr if the fit is done without including nebular emission. 

We have investigated the positions of our LAEs in the SFR versus stellar mass plane, and find that in general the z$\sim$4.5 LAE population follows a similar shape for the derived MS of galaxies at $z=$ 2, but the majority of our sample of LAEs lie above the z$\sim$2 MS of star forming galaxies as derived from Daddi et al.\ (2007), and the overall z$\sim$4 trend from Speagle et al.\ (2014).  Previous studies of higher redshift galaxies (e.g. Hathi et al.\ 2013 \& Bouwens et al.\ 2012) have also found that continuum--selected galaxies fall above the $z=$ 2 MS trend line, though not as high as our sample of LAEs.  We conclude that at a given stellar mass, both higher redshift galaxies, and galaxies exhibiting Ly$\alpha$ in emission, produce stars at a higher rate.  We do find a smaller subset of LAEs that fall below our trend line which are consistent with the $z=$ 2 MS results, indicating that there may be two mechanisms for Ly$\alpha$~escape.  One where the galaxies are young, massive, and undergoing a burst of star formation,which creates large amounts of Ly$\alpha$, and allows Ly$\alpha$~to more easily escape through an outflow in the galaxy.  In the second situation, the galaxies are older, more evolved, have lower SFRs, and have had time to create holes and locations in the galaxy for the Ly$\alpha$~to escape.  Although some fraction of these more evolved galaxies may truly be younger (due to degeneracies in the SED fitting process), it is possible that at least 15\% of our sample of LAEs are dramatically different than the typical LAE.  These possible scenarios of two methods of Ly$\alpha$ escape are directly testable by comparing the redshift of Ly$\alpha$ emission to the systemic redshift of a galaxy.

\acknowledgements  This work
is based in part on observations made with the \textit{Spitzer Space Telescope}, which is
operated by the Jet Propulsion Laboratory, California Institute of
Technology under a contract with NASA. Support for this work was
provided by NASA through an award issued by JPL/Caltech. Further
support for KDF and SLF was provided by the University of Texas at Austin. AD's research activities are supported by NOAO, which is operated by the Association of Universities for Research in Astronomy under a cooperative agreement with the US National Science Foundation; and in part by the Radcliffe Institute for Advanced Study at Harvard University.

\clearpage
\input paper_table2.tex

\input paper_table3.tex

\input paper_table4.tex

\input paper_table5.tex
\clearpage
\input paper_table6.tex
\input paper_table7.tex

\end{document}

%% file: paper_table1.tex
\begin{deluxetable*}{cccc}
\tablewidth{0pt}
\tablecaption{Summary of Source Coverage and Detections}
\tablehead{
\colhead{Sample} & \colhead{IRAC 3.6$\mu$m} & \colhead{IRAC 4.5$\mu$m} & \colhead{{\it HST} NIR at 1.1 $\&$ 1.6 $\mu$m}} 
\startdata
\hline
\multicolumn{4}{|c|}{Number of Sources with IRAC and {\it HST} NIR Coverage}\\
\hline
$z=$ 4.5 spec-z confirmed & 43 & 20 & 21 \\
$z=$ 4.5 candidates & 80 & 61 & 3 \\
$z=$ 5.7 spec-z confirmed & 5 & 3 & 2 \\
$z=$ 5.7 candidates & 7 & 7 & 1 \\
\hline
\multicolumn{4}{|c|}{Number of Sources with IRAC and {\it HST} NIR Detections}\\
\hline
$z=$ 4.5 spec-z$^{a}$ & 7 (12) & 4 (5) & 19 (20) \\
$z=$ 4.5 candidates $^{a}$ & 15 (18) & 7 (10) & 1 (2) \\
$z=$ 5.7 spec-z$^{a}$ & 2 (2) & 1 (1) & 1 (1) \\
$z=$ 5.7 candidates$^{a}$ & 6 (6) &  6 (6) & 1 (1) \\
\hline
$z=$ 4.5 Detection probabilities$^{b}$ & 18$\%$ (24$\%$) & 14$\%$ (19$\%$) & 83$\%$ (92$\%$) \\
$z=$ 5.7 Detection probabilities$^{b}$ & 67$\%$ (67$\%$) & 70$\%$ (70$\%$) & 67 $\%$ (67$\%$)
\enddata\\
\vspace{-2mm}

\tablecomments{Confirmed LAEs have spectroscopic followup. LAE
  candidates are based on combined narrow-band and broad-band
  photometry.  $^{a}$ Number of objects detected at greater than
  3$\sigma$ (> 2$\sigma$).  $^{b}$ Detection probabilities at
  3$\sigma$ (2$\sigma$) based on the total number of sources with IRAC
  or {\it HST} coverage in each category.}
\end{deluxetable*}

%% file: paper_table2.tex
\LongTables
\begin{landscape}
\begin{deluxetable*}{ccccccccccc}
\tabletypesize{\scriptsize}
\tablecaption{LAE z$\sim$ 4.5 Sample -- Flux Measurements}
\tablewidth{0pt}
\tablehead{\colhead{ID} & \colhead{RA (J2000)} & \colhead{Dec
    (J2000)} & \colhead{NB} & \colhead{R} & \colhead{I} &
\colhead{z'}& \colhead{F110W} & \colhead{F160W} & \colhead{IRAC 3.6} &
\colhead{IRAC 4.5}
}
\startdata
3c &  216.61862 &  35.635761 & 1.101 $\pm$ 0.087 & 0.139 $\pm$  0.016 &
0.099  $\pm$ 0.019 & 0.115  $\pm$ 0.142 & 0.091 $\pm$ 0.008 &  0.143 $\pm$ 0.011 & 0.340 $\pm$ 0.108 & -- \\
5c &  216.49904 &  35.586872 & 0.600 $\pm$  0.087 & 0.049 $\pm$ 0.017
&  0.059 $\pm$ 0.021  & -0.080 $\pm$ 0.148 & -- & -- & -0.018 $\pm$
0.508 & 0.524 $\pm$ 0.161 \\
7c  & 216.48054 &  35.510797 & 1.237 $\pm$  0.103 &  0.252 $\pm$  0.021 &
0.093  $\pm$ 0.029 &   0.040  $\pm$ 0.182 & -- & -- &  0.580  $\pm$ 0.200 &
0.209  $\pm$  0.064 \\
8c  & 216.42367  & 35.564161 & 0.758  $\pm$ 0.091 &  0.106
$\pm$ 0.016 &  0.088 $\pm$  0.020 &  0.117 $\pm$
0.141 & -- & -- & -- & 0.413  $\pm$  0.163 \\
10c &  216.21837 &  35.436978 & 0.431 $\pm$ 0.073 &  0.049 $\pm$ 0.016
& 0.039  $\pm$ 0.018 &  0.129 $\pm$ 0.152 & -- & -- & 0.270$\pm$
0.123 & -- \\
11c &  216.48675  & 35.703958 & 0.696 $\pm$ 0.079 & 0.121 $\pm$
0.018  & 0.140 $\pm$ 0.021  & 0.149 $\pm$  0.156 &  0.156 $\pm$
0.006 &   0.115 $\pm$ 0.008 & 0.192  $\pm$  0.132 & -- \\
12c  & 216.41458 &  35.650358 & 0.492 $\pm$ 0.079 & 0.084 $\pm$ 0.016
& 0.111 $\pm$  0.020 & 0.135 $\pm$ 0.143 & -- & -- & -- &  0.737 $\pm$ 0.240 \\
16c &  216.30292  & 35.631914 & 0.913 $\pm$  0.080 & 0.155 $\pm$ 0.017 &  0.116 $\pm$ 0.019 &  0.076
$\pm$ 0.147 & 0.219 $\pm$ 0.006 & 0.205 $\pm$ 0.009 &  0.28 $\pm$
0.196 & 0.263 $\pm$ 0.298 \\
17c & 216.29983 & 35.653361 & 0.457 $\pm$ 0.080 & 0.067 $\pm$  0.017
&  0.093 $\pm$ 0.021 & 0.216  $\pm$ 0.147 & 0.027 $\pm$ 0.006  &
0.010  $\pm$  0.010 & 0.174 $\pm$  0.065 & 0.166  $\pm$ 0.301 \\
18c & 216.27658 & 35.638619 & 1.110 $\pm$ 0.086 & 0.169 $\pm$ 0.017
& 0.188 $\pm$ 0.020 & 0.136 $\pm$ 0.153 & 0.141 $\pm$ 0.006 & 0.142
$\pm$ 0.009 & 0.452 $\pm$ 0.352 & 0.258 $\pm$ 0.257 \\
21c &  216.42721 & 35.440425 & 0.431 $\pm$ 0.080 & 0.086 $\pm$ 0.015 &   0.086
$\pm$ 0.019 &  0.198 $\pm$ 0.143 & 0.111 $\pm$ 0.007 &  0.075 $\pm$ 0.008 & -0.448 $\pm$ 0.141 & -- \\
22c  & 216.38225  & 35.447653 & 0.3716 $\pm$   0.082 &  0.037 $\pm$  0.015 &
0.085  $\pm$  0.020 &  -0.043 $\pm$   0.144 & 0.022 $\pm$ 0.005  &  -0.046 $\pm$   0.044 &
-0.260 $\pm$   0.140 & -- \\
23c &  216.35883  &  35.425083 & 0.346 $\pm$ 0.077 & 0.047 $\pm$ 0.016 & 0.086
$\pm$ 0.020 &  -0.046  $\pm$ 0.147 & 0.027 $\pm$ 0.005 & 0.031 $\pm$
0.008 & 0.072 $\pm$ 0.219 & -- \\
27c  & 216.77608  &  35.539933 & 0.427 $\pm$  0.0626 & 0.103 $\pm$
0.017 & 0.044 $\pm$ 0.021 &  0.220 $\pm$  0.160 & -- & -- & 0.148 $\pm$ 0.073 &   -0.214 $\pm$ 0.135 \\
29c  &  216.72292 &  35.565436 & 0.541 $\pm$ 0.059 & 0.092 $\pm$ 0.016
& 0.089 $\pm$ 0.020 &  -0.047 $\pm$ 0.146 & 0.116 $\pm$ 0.009  &
0.138 $\pm$ 0.010 & -0.161 $\pm$ 0.122 &  0.231  $\pm$ 0.141 \\
30c  & 216.61433 & 35.621208 & 0.741 $\pm$ 0.064 & 0.189 $\pm$ 0.017 &
0.265 $\pm$ 0.022 & 0.192 $\pm$ 0.146 & -- & -- &  0.519 $\pm$ 0.140 & -- \\
32c  & 216.38700 & 35.503331 & 1.864 $\pm$  0.164 & 0.236 $\pm$  0.020 &
0.229  $\pm$  0.024 & 0.204 $\pm$ 0.152 & 0.182 $\pm$  0.012 &
0.129 $\pm$ 0.013 & -0.025 $\pm$ 0.370  & 0.040  $\pm$   3.22 \\
35c &  216.28437 & 35.664389 & 0.544 $\pm$  0.066 & 0.074 $\pm$
0.019 &   0.079 $\pm$ 0.024 &   -0.101  $\pm$ 0.167 & 0.0433  $\pm$ 0.006 & 0.028 $\pm$   0.009 & -0.183 $\pm$ 0.169 &
-0.125 $\pm$ 0.176 \\
37c & 216.25687  & 35.614272 & 0.963 $\pm$ 0.062 & 0.147 $\pm$ 0.016 & 0.142
$\pm$ 0.019 & 0.158 $\pm$ 0.147 & 0.082 $\pm$ 0.009 & 0.103 $\pm$
0.011 & 0.313 $\pm$ 0.178 & -- \\
40c & 216.20854  & 35.499861 & 0.875 $\pm$ 0.061 & 0.082 $\pm$ 0.016 & 0.143
$\pm$ 0.020 & 0.220  $\pm$ 0.154 & 0.158 $\pm$ 0.006 &  0.150 $\pm$ 0.008 & 0.112 $\pm$ 0.154 & -- \\
41c & 216.18908 & 35.482889 & 0.416 $\pm$ 0.058 & 0.191 $\pm$ 0.017 & 0.249 $\pm$ 0.020 &
0.296  $\pm$ 0.155 & 0.337 $\pm$ 0.006 &  0.310 $\pm$ 0.009 & 0.678 $\pm$ 0.106 & -- \\
42c & 216.18821 & 35.488669 & 0.342 $\pm$ 0.058 & 0.085  $\pm$ 0.016 & 0.216
$\pm$ 0.020 & 0.351 $\pm$ 0.156 & 0.283 $\pm$ 0.005 & 0.247 
$\pm$ 0.007 & 0.487  $\pm$ 0.063 & -- \\
43c & 216.45162 & 35.461061 & 0.918 $\pm$ 0.148 & 0.081 $\pm$ 0.016 &
0.102 $\pm$ 0.020 & 0.262 $\pm$ 0.143 & -- & -- & 0.613 $\pm$ 0.102 &  -0.260
$\pm$ 0.622 \\
44c  & 216.44921 & 35.699917 & 0.819 $\pm$ 0.145 & 0.105 $\pm$ 0.018 &  0.033 $\pm$ 0.021
& 0.077  $\pm$ 0.153 & 0.055 $\pm$ 0.006 &  0.017 $\pm$ 0.003 &  0.289
$\pm$ 0.125 & -- \\
45c & 216.43546 & 35.723436 & 0.254 $\pm$ 0.057 & 0.012 $\pm$ 0.016 & 0.036 
$\pm$ 0.020 & 0.204 $\pm$ 0.150 & 0.073 $\pm$ 0.010 & 0.086 $\pm$ 0.013 & 0.417 $\pm$ 0.138 & -- \\
46c & 216.15983 & 35.394056 & 0.559 $\pm$ 0.063 & 0.101 $\pm$ 0.017 & 0.172 $\pm$ 0.022
&  0.503  $\pm$ 0.169 & -- & -- & -- & 0.545 $\pm$ 0.211 \\
48c & 216.48629 & 35.709244 & 0.323 $\pm$ 0.082 & 0.019 $\pm$ 0.016 & 0.027 $\pm$ 0.019 &
-0.068 $\pm$ 0.156 & 0.025 $\pm$ 0.005 &  0.026 $\pm$ 0.008 & 0.028  $\pm$ 0.863 & -- \\
49c & 216.42500 & 35.432442 & 0.457 $\pm$ 0.087 & 0.021 $\pm$ 0.016 &  0.103  $\pm$ 0.020
& -0.182 $\pm$ 0.148 & 0.149 $\pm$ 0.012 &  0.207  $\pm$ 0.013 & -0.182  $\pm$ 0.151 & -- \\
53c & 216.78825 & 35.402408 & 0.289 $\pm$ 0.060 & 0.103 $\pm$ 0.017 & 0.083 $\pm$ 0.021 &
0.088 $\pm$ 0.152 & -- & -- & 0.487 $\pm$ 0.159 & -- \\
56c & 216.47771 & 35.555628 & 0.306 $\pm$ 0.065 & 0.092 $\pm$ 0.018 &  0.251 $\pm$ 0.022 
&  0.693 $\pm$ 0.161 & -- & -- & -- & 4.607  $\pm$ 0.169 \\

10nc & 216.65613 & 35.289208 & 0.389 $\pm$  0.073 &  0.045 $\pm$ 0.016 & 0.051
$\pm$ 0.020 &  0.128  $\pm$ 0.148 & -- & -- & 0.474  $\pm$ 0.164 & --\\
22nc  & 216.23679 & 35.698278 & 0.835 $\pm$ 0.078 & 0.183 $\pm$ 0.017 & 0.243 $\pm$ 0.021
&  0.029  $\pm$ 0.164 & -- & -- & 0.359 $\pm$ 0.080 & 0.420  $\pm$ 0.221 \\
23nc & 216.22208 & 35.638567 & 0.420 $\pm$ 0.088 & 0.100 $\pm$ 0.016 & 0.127 $\pm$ 0.019 &
0.156 $\pm$ 0.150 & -- & -- & 0.072 $\pm$ 0.219 & 0.355 $\pm$ 0.329 \\
26nc  & 216.13592  & 35.390422 & 0.422 $\pm$ 0.108 & 0.130 $\pm$ 0.020 & 0.039
$\pm$ 0.024 & -0.026 $\pm$ 0.188 & -- & -- & -- & 0.331 $\pm$ 0.157 \\
31nc & 216.78079 & 35.422856 & 0.363 $\pm$ 0.073 & 0.199 $\pm$ 0.017 & 0.281 $\pm$ 0.021
& 0.156  $\pm$ 0.155 & -- & -- & 0.936 $\pm$ 0.129 & -- \\
36nc & 216.59233 & 35.570203 & 0.417 $\pm$ 0.073 & 0.059 $\pm$ 0.016 & 0.093 $\pm$ 0.020 &
0.008  $\pm$ 0.143 & -- & -- & 0.369 $\pm$ 0.138 & -- \\
40nc & 216.41279  & 35.755600 & 0.399 $\pm$ 0.079 & 0.090  $\pm$ 0.016 & 0.098 $\pm$ 0.020
&  0.076 $\pm$ 0.151 & -- & -- & 0.313 $\pm$ 0.081 & -- \\
43nc  & 216.31254 & 35.675472 & 0.441 $\pm$ 0.086 & 0.040 $\pm$ 0.016 & 0.053 $\pm$ 0.019
& 0.752 $\pm$ 0.615 & -- & -- & 0.256 $\pm$ 0.161 & 0.537 $\pm$ 0.190 \\
44nc & 216.28917 & 35.749864 & 0.340 $\pm$ 0.076 & 0.072 $\pm$ 0.017 & 0.152 $\pm$ 0.021 &
0.216  $\pm$ 0.154 & -- & -- & -- & 0.380  $\pm$ 0.146 \\
57nc & 216.66546  & 35.505661 & 0.439 $\pm$ 0.082 & 0.036 $\pm$ 0.017 & 0.041 $\pm$ 0.022
& -0.104 $\pm$ 0.153 & -- & -- & 0.180 $\pm$ 0.189 & 0.359 $\pm$ 0.174\\
58nc & 216.65525 & 35.736069 &  0.957 $\pm$ 0.062 & 0.125 $\pm$ 0.017 & 0.197 $\pm$ 0.020
&  0.188 $\pm$ 0.142 & -- & -- & -- & 0.322  $\pm$ 0.090 \\
60nc & 216.61350 & 35.350981 & 0.658 $\pm$ 0.060 & 0.194 $\pm$ 0.016 & 0.226 $\pm$ 0.019 &
0.245 $\pm$ 0.148 & -- & -- & 1.085 $\pm$ 0.220 & -- \\
62nc  & 216.58488 &  35.256228 & 0.932 $\pm$ 0.064 & 0.101 $\pm$ 0.016 & 0.063
$\pm$ 0.021 & 0.172 $\pm$ 0.153 & -- & -- & 0.448 $\pm$ 0.138 & -- \\
66nc & 216.46129 & 35.762097 & 0.364 $\pm$ 0.081 & 0.236 $\pm$ 0.018 & 0.272 $\pm$ 0.021
& 0.645 $\pm$ 0.158 & -- & -- & 0.205 $\pm$ 0.057 & -- \\
76nc & 216.82129 & 35.441064 & 0.445 $\pm$ 0.064 & 0.120 $\pm$ 0.020 & 0.187 $\pm$ 0.025 &
0.117 $\pm$ 0.159 & -- & -- & 1.085 $\pm$ 0.110 & -- \\
79nc & 216.78329 & 35.353919 & 1.056 $\pm$ 0.064 & 0.228 $\pm$ 0.016 & 0.279 $\pm$ 0.021 &
0.618 $\pm$ 0.162 & -- & -- & -- & 0.870  $\pm$ 0.152 \\
83nc & 216.67629 & 35.493894 & 0.342 $\pm$ 0.061 & 0.017 $\pm$ 0.016 & -0.009  $\pm$ 0.021
& 0.082 $\pm$ 0.146 & -- & -- & 0.461 $\pm$ 0.146  &  -0.165 $\pm$ 0.155 \\
87nc & 216.64671 & 35.252078 & 0.385 $\pm$ 0.062  & 0.111 $\pm$ 0.016 & 0.149 $\pm$ 0.021
&  0.255 $\pm$ 0.147 & -- & -- & 0.249 $\pm$ 0.086 & -- \\
88nc & 216.64079 & 35.450736 & 0.507 $\pm$ 0.060 & 0.285 $\pm$ 0.017 & 0.461 $\pm$ 0.021
& 0.890 $\pm$ 0.146 & -- & -- & 2.603 $\pm$ 0.072  &  2.125 $\pm$ 0.117 \\
90nc & 216.63858 & 35.439358 &  0.455 $\pm$ 0.062 & 0.032 $\pm$ 0.017 & -0.010 $\pm$ 0.021
& 0.061 $\pm$ 0.154 & -- & -- & 0.323 $\pm$ 0.301 &  0.395  $\pm$ 0.294 \\
96nc & 216.58721 & 35.580183 & 0.407 $\pm$ 0.060 & 0.043 $\pm$  0.016 & 0.145 $\pm$ 0.019
& 0.266 $\pm$ 0.142 & -- & -- & 0.302 $\pm$ 0.0840 & -- \\
98nc & 216.57875 & 35.287997 & 1.067 $\pm$ 0.065 & 0.185 $\pm$ 0.017 & 0.226 $\pm$ 0.020
& -0.257 $\pm$ 0.153 & -- & -- & 0.491 $\pm$ 0.160 & -- \\
104nc & 216.43054 & 35.709911 & 0.444 $\pm$  0.058 & 0.123 $\pm$ 0.016 &
0.119 $\pm$ 0.020 & 0.125 $\pm$ 0.148 & -- & -- & 0.340 $\pm$ 0.085 & -- \\
110nc & 216.27879 & 35.817781 & 0.372 $\pm$ 0.074 & 0.067 $\pm$ 0.018 & 0.048
$\pm$ 0.023 &  -0.081 $\pm$ 0.194 & -- & -- & -- & 3.976 $\pm$ 0.255 \\
111nc & 216.27292 & 35.772394 & 0.336 $\pm$ 0.059 & 0.018 $\pm$ 0.014 & -0.029 $\pm$ 0.018
& -0.050 $\pm$ 0.149 & -- & -- & -- & 0.387 $\pm$ 0.145 \\
114nc & 216.24808 & 35.606881 & 0.615 $\pm$ 0.062 & 0.209 $\pm$ 0.017 & 0.376 $\pm$ 0.019
& 1.006 $\pm$ 0.149 & 0.275 $\pm$ 0.020 & 0.453 $\pm$ 0.025 & 5.800 $\pm$ 0.159 & -- \\
121nc & 216.77842 & 35.574064 & 0.795 $\pm$ 0.148 & 0.090 $\pm$ 0.018 & 0.224 $\pm$ 0.021
& 0.551 $\pm$ 0.155 & -- & -- & 0.630 $\pm$ 0.157 & 0.666 $\pm$ 0.135 \\
138nc & 216.20413 & 35.333586 & 0.702 $\pm$ 0.149 & 0.113 $\pm$ 0.017 & 0.190 $\pm$ 0.022
& 0.402 $\pm$ 0.154 & -- & -- & -- & 0.854 $\pm$ 0.149 \\
182nc & 216.66083 & 35.775336 & 0.840 $\pm$ 0.061 & 0.625 $\pm$ 0.018 & 0.801 $\pm$ 0.022
& 1.124 $\pm$ 0.151 & -- & -- & -- & 1.642 $\pm$ 0.242 \\
191nc & 216.75946 & 35.425897 & 0.616 $\pm$ 0.084 & 0.077 $\pm$ 0.016 & 0.064
$\pm$ 0.020 & 0.264  $\pm$ 0.156 & -- & -- & 0.575 $\pm$ 0.260 & -- \\
196nc & 216.21612 & 35.553250 & 0.396 $\pm$ 0.060 & 0.064 $\pm$ 0.025 & 0.102 $\pm$ 0.050
& 0.150 $\pm$ 0.231 & -- & -- & 0.510 $\pm$ 0.061 & -- 
\enddata
\tablecomments{Source IDs ending in 'c' are LAEs that have been
  spectroscopically confirmed. Source IDs ending in 'nc' have a
  photometric redshift based on narrow-band imaging, and have not been
targeted spectroscopically. RA, Dec in degrees. Fluxes are all given
in $\mu$Jy. No coverage in a {\it Spitzer} or {\it HST} band is
denoted by -- .}
\end{deluxetable*}
\clearpage
\end{landscape}

%% file: paper_table3.tex
\begin{landscape}
\begin{deluxetable*}{ccccccccccc}
\tabletypesize{\scriptsize}
\tablecaption{LAE z$\sim$ 5.7 Sample -- Flux Measurements}
\tablewidth{0pt}
\tablehead{\colhead{ID} & \colhead{RA (J2000)} & \colhead{Dec
    (J2000)} & \colhead{NB} & \colhead{R} & \colhead{I} &
\colhead{z'}& \colhead{F110W} & \colhead{F160W} & \colhead{IRAC 3.6} &
\colhead{IRAC 4.5}
}
\startdata
54c &  216.69638 &  35.603511 & 1.278 $\pm$ 0.122 & -0.026 $\pm$  0.019 &
0.081  $\pm$ 0.023 & 0.499  $\pm$ 0.151 & -- &  -- & -1.056 $\pm$ 0.281 & -- \\
55c &   216.62629 &  35.672933 & 0.784 $\pm$ 0.119 & -0.003 $\pm$ 0.018  & 0.044 $\pm$ 0.022 &  -0.060 $\pm$ 0.155 & -- & -- & 1.136 $\pm$ 0.273 & 0.134 $\pm$ 2.107 \\
57c & 216.39138  &  35.506383 & 1.170 $\pm$ 0.136 & 0.057 $\pm$  0.028 &  0.662 $\pm$  0.062 & 1.438 $\pm$ 0.185 & 2.134 $\pm$ 0.012   &  3.026  $\pm$ 0.014 &  3.277 $\pm$ 0.210 & 1.817  $\pm$ 0.233 \\
58c & 216.38038  & 35.427475 & 0.715 $\pm$ 0.115 & 0.025 $\pm$ 0.017 &
0.023 $\pm$ 0.020 & 0.024 $\pm$ 0.1465 & 0.006 $\pm$ 0.005  &  0.011
$\pm$  0.008 & -0.533 $\pm$  0.139 & -- \\
63c & 216.40554 &  35.480000 & 0.631 $\pm$ 0.102 & -0.035 $\pm$ 0.016 & 0.052 $\pm$ 0.021 &  -0.194 $\pm$ 0.144 & -- & -- &  -0.375  $\pm$ 0.140 & 0.121 $\pm$ 0.209 \\
202nc & 216.73854 & 35.615664 & 0.563 $\pm$ 0.109 & -0.006 $\pm$ 0.016 &  0.177 $\pm$  0.021 &0.408 $\pm$ 0.148 & -- & -- & 0.886 $\pm$ 0.22 & 1.379 $\pm$  0.371 \\
203nc & 216.63029 & 35.496514 & 0.793 $\pm$  0.111 & -0.008 $\pm$ 0.015 & 0.049 $\pm$ 0.020 &  -0.077 $\pm$ 0.145 & -- & -- &  -0.037  $\pm$ 0.134 &  -0.116  $\pm$ 0.188 \\
207nc & 216.55396 &  35.640358 & 1.247 $\pm$ 0.116 & 0.061 $\pm$ 0.016 & 0.565 $\pm$ 0.021  & 1.774 $\pm$ 0.150 & -- & -- &  4.523 $\pm$ 0.124 & -- \\
209nc & 216.40683 &  35.637464 & 0.799 $\pm$ 0.114 & 0.123 $\pm$  0.017 & 0.316$\pm$ 0.022 & 0.517 $\pm$ 0.148 & -- & -- & -- & 11.573   $\pm$ 1.062 \\
210nc & 216.36638  & 35.727917 & 1.117 $\pm$ 0.117 & 0.140 $\pm$ 0.016 &  0.443 $\pm$ 0.020 &  1.133 $\pm$ 0.148 & -- & -- & -- & 17.356 $\pm$ 0.477 \\
222nc & 216.72350  & 35.419058 & 0.680 $\pm$ 0.097 & 0.045 $\pm$ 0.016 & 0.129 $\pm$ 0.021 & -0.014 $\pm$ 0.154 & -- & -- & 23.092 $\pm$ 2.118 & 23.521 $\pm$ 1.509 \\
223nc & 216.72079  & 35.572011 & 0.565 $\pm$ 0.091 & -0.003 $\pm$ 0.015 & 0.024 $\pm$ 0.020 & 0.152 $\pm$ 0.148 & 0.184 $\pm$  0.012 & 0.180 $\pm$ 0.015 & 0.334 $\pm$ 0.068 & 0.380 $\pm$ 0.095 \\
224nc & 216.63808 & 35.313544 & 0.542 $\pm$ 0.096 & 0.172 $\pm$ 0.017 & 0.216 $\pm$ 0.020 & 0.328 $\pm$ 0.149 & -- & -- & 2.726 $\pm$ 0.125 & 3.369 $\pm$  0.309 \\
231nc & 216.46479 & 35.731386 & 0.796 $\pm$  0.097 & 0.146 $\pm$ 0.017
& 0.343 $\pm$ 0.021 & 0.519 $\pm$ 0.153 & -- & -- & 0.474 $\pm$ 0.109
& -- 
\enddata
\tablecomments{Source IDs ending in 'c' are LAEs that have been
  spectroscopically confirmed. Source IDs ending in 'nc' have a
  photometric redshift based on narrow-band imaging, and have not been
targeted spectroscopically. The observed narrow-band for the $z=$5.7
sources is at 815 or 823 nm. RA, Dec in degrees. Fluxes are all given
in $\mu$Jy. No coverage in a {\it Spitzer} or {\it HST} band is denoted by --}
\end{deluxetable*}
\clearpage
\end{landscape}

%% file: paper_table4.tex
\vspace{-0.2cm}
\begin{deluxetable*}{cccccccccc}
\tabletypesize{\footnotesize}
\tablecaption{$z=$4.5 IRAC Detected LAEs -- Model SED Fitting Results}
\tablewidth{0pt}
\tablehead{
\colhead{ID} & \colhead {Reduced $\chi^{2}$} & \colhead{Mass} &
\colhead{Mass} & \colhead{Age} &
\colhead{Age} & \colhead{E(B-V)} &
\colhead{E(B-V)} & \colhead{Z} & \colhead{Z} \\
\colhead{} & \colhead{of Best Fit} & \colhead{Best Fit} &
\colhead{68$\%$ Range} & \colhead{Best Fit} &
\colhead{68$\%$ Range} & \colhead{Best Fit} &
\colhead{68$\%$ Range} & \colhead{Best Fit} & \colhead{68$\%$ Range} \\
\colhead{} & \colhead{} &\colhead{(10$^{8}$ M$_{\odot}$)} &
\colhead{(10$^{8}$ M$_{\odot}$)} & \colhead{(Myr)} &
\colhead{(Myr)} & \colhead{(mag)} &
\colhead{(mag)}& \colhead{(Z$_{\odot}$)} & \colhead{(Z$_{\odot}$)}
}
\startdata
3c & 9.6 & 70 & 2.0 -- 5.9 & 1015 & 10 -- 10 & 0.03 & 0.05 -- 0.19 & 0.02  & 0.02 -- 0.02 \\
7c & 30.2 & 1.5 & 1.4 -- 7.7 & 10 & 10 -- 10 & 0.00 & 0.00 -- 0.13 & 0.20 & 0.02 -- 0.20 \\
12c & 0.5 &168 & 54 -- 204 & 1015 & 10 -- 1015 & 0.12 & 0.04 -- 0.39 & 0.02 & 0.02 -- 0.20 \\
30c & 10.4 & 61 & 17 -- 109 & 806 & 45 -- 1015 & 0.00 & 0.00 -- 0.02 & 0.20& 0.02 -- 0.40 \\
41c & 3.1 & 12 & 11 -- 21 & 10 & 10 -- 20 & 0.18 & 0.15 -- 0.21 & 1.00 & 1.00 -- 1.00 \\
42c & 4.9 & 49 & 47 -- 74 & 57 & 10 -- 72 & 0.00 & 0.00 -- 0.30 & 1.00 & 0.02 -- 1.00 \\
43c & 2.3 & 20 & 15 -- 35 & 10 & 10 -- 12 & 0.33 & 0.27 -- 0.36 & 0.02 & 0.02 -- 0.02\\
45c & 1.9 & 71 & 36 -- 208 & 102 & 10 -- 570 & 0.16 & 0.00 -- 0.60 & 0.20 & 0.02 -- 1.00 \\
53c & 4.4 & 11 & 0.8 -- 22 & 10 & 10 -- 10 & 0.27 & 0.00 -- 0.33 & 0.02 & 0.02 -- 0.02 \\
56c & 5.0 & 1184 & 973 -- 1954 & 18 & 10 -- 90 & 0.60 & 0.24 -- 0.66 & 0.02 & 0.02 -- 0.4 \\
22nc & 2.0 & 46.6 & 7.3 -- 55.8 & 806 & 10 -- 1015 & 0.00 & 0.00 -- 0.09 & 0.02 & 0.02 -- 0.40 \\
31nc & 0.9 & 178 & 36 -- 200 & 1015 & 20 -- 1015 & 0.02 & 0.01 -- 0.21 & 0.20 & 0.02 -- 0.40 \\
40nc & 0.5 & 74 & 5.2 -- 28.3 & 10 & 10 -- 45 & 0.24 & 0.11 -- 0.27 & 0.20 & 0.02 -- 0.20 \\
58nc & 0.01 & 13.8 & 12.2 -- 52.5 & 19 & 10 -- 905 & 0.19 & 0.00 -- 0.27 & 0.40 & 0.02 -- 0.40 \\
60nc & 3.3 & 24.9 & 17.8 -- 130 & 10 & 10 -- 1015 & 0.27 & 0.07 -- 0.30 & 0.02 & 0.02 -- 0.20 \\
62nc & 6.9 & 10.3 & 3.3 -- 17.1 & 10 & 10 -- 10 & 0.27 & 0.14 -- 0.33 & 0.02 & 0.02 -- 0.02 \\
66nc & 6.1 & 4.5 & 2.7 -- 8.4 & 13 & 10 -- 20 & 0.05 & 0.01 -- 0.08 & 0.02 & 0.02 -- 0.02 \\
76nc & 0.3 & 270 & 30 -- 278 & 1015 & 10 -- 1015 & 0.10 & 0.08 -- 0.33 & 0.02 & 0.02 -- 0.40 \\
79nc & 6.8 & 167 & 60.5 -- 185 & 1015 & 10 -- 1015 & 0.02 & 0.00 -- 0.36 & 0.02 & 0.02 -- 0.02 \\
83nc & 7.3 & 33.7 & 16.2 -- 54.2 & 10 & 10 -- 10 & 0.60 & 0.42 -- 0.75 & 0.02 & 0.02 -- 0.02  \\
88nc & 7.8 & 160 & 109 -- 317 & 57 & 20 -- 102 & 0.27 & 0.27 -- 0.30 & 0.20 & 0.02 -- 1.00 \\
96nc & 5.8 & 34.6 & 26 -- 55.7 & 64 & 45 -- 102 & 0.00 & 0.00 -- 0.11 & 1.00 & 0.02 -- 1.00 \\
98nc & 10.4 & 8.7 & 3.4 -- 42.8 & 10 & 10 -- 719 & 0.17 & 0.00 -- 0.19 & 0.02 & 0.02 -- 0.20 \\
104nc & 1.7 & 6.4 & 3.7 -- 10.3 & 10 & 10 -- 10 & 0.19 & 0.10 -- 0.21 & 0.02 & 0.02 -- 0.02 \\
110nc & 6.5 & 1187 & 1053 -- 1381 & 570 & 570 -- 570 & 0.05 & 0.01 -- 0.09 & 1.00 & 1.00 -- 1.00 \\
114nc & 40.6 & 2082 & 2020 -- 2561 & 1015 & 1015 -- 1015 & 0.27 & 0.27 -- 0.30 & 0.20 & 0.20 -- 0.20 \\
121nc & 8.3 & 77.9 & 59.2 -- 124 & 72 & 10 -- 80 & 0.04 & 0.00 -- 0.36 & 1.00 & 0.02 -- 1.00 \\
138nc & 1.6 & 113 & 86.0 -- 205 & 10 & 10 -- 1015 & 0.45 & 0.36 -- 0.51 & 0.02 & 0.02 -- 0.20 \\
182nc & 8.4 & 268 & 177 -- 315 & 806 & 806 -- 1015 & 0.00 & 0.00 -- 0.00 & 0.02 & 0.02 -- 0.20 \\
196nc & 0.1 & 130 & 75.4 -- 116 & 1015 & 10 -- 1015 & 0.08 & 0.02 -- 0.33 & 0.40& 0.02 -- 1.00 
\enddata
\end{deluxetable*}

%% file: paper_table5.tex
\vspace{-0.2cm}
\begin{deluxetable*}{cccccccccc}
\tabletypesize{\footnotesize}
\tablecaption{$z=$4.5 {\it HST} NIR Detected LAEs with IRAC Upper Limits -- Model SED Fitting Results}
\tablewidth{0pt}
\tablehead{
\colhead{ID} & \colhead {Reduced $\chi^{2}$} & \colhead{Mass} &
\colhead{Mass} & \colhead{Age} &
\colhead{Age} & \colhead{E(B-V)} &
\colhead{E(B-V)} & \colhead{Z} & \colhead{Z} \\
\colhead{} & \colhead{of Best Fit} & \colhead{Best Fit} &
\colhead{68$\%$ Range} & \colhead{Best Fit} &
\colhead{68$\%$ Range} & \colhead{Best Fit} &
\colhead{68$\%$ Range} & \colhead{Best Fit} & \colhead{68$\%$ Range} \\
\colhead{} & \colhead{} &\colhead{(10$^{8}$ M$_{\odot}$)} &
\colhead{(10$^{8}$ M$_{\odot}$)} & \colhead{(Myr)} &
\colhead{(Myr)} & \colhead{(mag)} &
\colhead{(mag)}& \colhead{(Z$_{\odot}$)} & \colhead{(Z$_{\odot}$)}
}
\startdata
11c & 2.1 &2.0 & 2.7 -- 10.3 & 10 & 10 -- 20 & 0.06 & 0.04 -- 0.18 & 1.00 & 0.02 -- 1.00 \\
16c & 5.1 & 7.4 & 6.9 -- 14.8 & 10 & 10 -- 10 & 0.18 & 0.16 -- 0.24 & 1.00 & 1.00 -- 1.00 \\
17c & 5.2 & 0.33 & 0.3 -- 16.3 & 10 & 10 -- 1015 & 0.00 & 0.00 -- 0.00 & 0.02 & 0.02 -- 0.02\\
18c & 1.2 & 2.5 & 2.1 -- 23.5 & 10 & 10 -- 905 & 0.07 & 0.00 -- 0.09 & 0.20 & 0.02 -- 0.20 \\
21c & 8.5 & 3.1 & 1.3 -- 5.5 & 10 & 10 -- 10 & 0.06 & 0.01 -- 0.13 & 0.02 & 0.02 -- 1.00 \\
22c & 9.4 & 0.3 & 0.2 -- 0.6 & 11 & 10 -- 12 & 0.00 & 0.00 -- 0.00 & 0.02 & 0.02 -- 0.02 \\
23c & 3.4 & 0.3 & 0.3 -- 0.6 & 10 & 10 -- 12 & 0.00 & 0.00 -- 0.00 & 0.20 & 0.02 -- 0.02 \\
29c & 5.0 & 10.3 & 2.6 -- 10.4 & 18 & 10 -- 18 & 0.07 & 0.01 -- 0.15 & 1.00 & 0.02 -- 1.00 \\
32c & 0.5 & 1.7 & 1.6 -- 2.1 & 10 & 10 -- 11 & 0.00 & 0.00 -- 0.03 & 0.02 & 0.02 -- 0.02 \\
35c & 1.5 & 0.4 & 0.4 -- 0.6 & 10 & 10 -- 11 & 0.00 & 0.00 -- 0.00 & 0.02 & 0.02 -- 0.02 \\
37c & 7.5 & 1.0 & 0.9 -- 1.1 & 10 & 10 -- 10 & 0.09 & 0.00 -- 0.01 & 0.20 & 0.02 -- 0.20 \\
40c & 0.7 & 13.8 & 5.9 -- 25.1 & 10 & 10 -- 40 & 0.19 & 0.04 -- 0.21 & 0.02 & 0.02 -- 1.00 \\
44c & 17. 0 & 0.4 & 0.3 -- 0.4 & 10 & 10 -- 10 & 0.00 & 0.00 -- 0.00 & 0.02 & 0.02 -- 0.02 \\
48c & 0.4 & 1.7 & 0.4 -- 60.6 & 10 & 10 -- 404 & 0.24 & 0.00 -- 0.30 & 0.02 & 0.02 -- 1.00 \\
49c & 7.0 & 65.8 & 39.5 -- 67.0 & 10 & 10 -- 10 & 0.39 & 0.33 -- 0.39 & 0.02 & 0.02 -- 0.02 
\enddata
\end{deluxetable*}

%% file: paper_table6.tex
\vspace{-0.2cm}
\begin{deluxetable*}{cccccccccc}
\tabletypesize{\footnotesize}
\tablecaption{$z=$5.7 IRAC Detected LAEs -- Model SED Fitting Results}
\tablewidth{0pt}
\tablehead{
\colhead{ID} & \colhead {Reduced $\chi^{2}$} & \colhead{Mass} &
\colhead{Mass} & \colhead{Age} &
\colhead{Age} & \colhead{E(B-V)} &
\colhead{E(B-V)} & \colhead{Z} & \colhead{Z} \\
\colhead{} & \colhead{of Best Fit} & \colhead{Best Fit} &
\colhead{68$\%$ Range} & \colhead{Best Fit} &
\colhead{68$\%$ Range} & \colhead{Best Fit} &
\colhead{68$\%$ Range} & \colhead{Best Fit} & \colhead{68$\%$ Range} \\
\colhead{} & \colhead{} &\colhead{(10$^{8}$ M$_{\odot}$)} &
\colhead{(10$^{8}$ M$_{\odot}$)} & \colhead{(Myr)} &
\colhead{(Myr)} & \colhead{(mag)} &
\colhead{(mag)}& \colhead{(Z$_{\odot}$)} & \colhead{(Z$_{\odot}$)}
}
\startdata
55c & 1.1 &  48.8 & 38.3 -- 505 & 1 & 1 -- 404 & 0.45 & 0.00 -- 0.51 & 0.20 &
0.20 -- 1.00 \\
57c & 27.5 & 497 & 486 -- 636 & 10 & 10 -- 10 & 0.24 & 0.24 -- 0.27 & 0.02 &
0.02 -- 0.02 \\
202nc & 2.2 & 173 & 31.2 -- 175 & 7 & 2 -- 57 & 0.33 & 0.00 -- 0.36 & 0.02 &
0.02 -- 1.00 \\
207nc & 2.1 & 760 & 706 -- 859 & 72 & 72 -- 102 & 0.00 & 0.00 -- 0.03 & 1.00
& 0.20 -- 1.00 \\
209nc & 24.2 & 5150 & 542 -- 5339 & 509 & 2 -- 509 & 0.00 & 0.00 -- 0.42 &
1.00 & 0.40 -- 1.00 \\ 
210nc & 28.7 & 5764 & 5621 -- 7434 & 509 & 509 -- 509 & 0.00 & 0.00 -- 0.06 & 0.40 &
0.02 -- 1.00 \\
223nc & 1.5 & 53.2 & 42.3 -- 71.6 & 57 & 50 -- 81 & 0.00 & 0.00 -- 0.00 &
1.00 & 0.20 -- 1.00 \\
224nc & 61.4 & 1139 & 86.2 -- 1332 & 806 & 3 -- 806 & 0.15 & 0.03 -- 0.30 &
0.20 & 0.02 -- 0.20 \\
231nc & 29.0 & 9.95 & 2.89 -- 35.6 & 7 & 3 -- 19 & 0.00 & 0.00 -- 0.00 & 0.20
& 0.02 -- 0.40 
\enddata
\end{deluxetable*}

%% file: paper_table7.tex
\vspace{-0.2cm}
\begin{deluxetable*}{cccccccc}
\tabletypesize{\footnotesize}
\tablecaption{$z=$ 4.5 LAE Stacks -- Model SED Fitting Results}
\tablewidth{0pt}
\tablehead{
\colhead{Stack}& \colhead{Reduced $\chi^{2}$} & \colhead{Mass} &
\colhead{Mass} & \colhead{Age} &
\colhead{Age} & \colhead{E(B-V)} &
\colhead{E(B-V)} \\
\colhead{}& \colhead{of Best Fit}& \colhead{Best Fit}&
\colhead{68$\%$ Range} & \colhead{Best Fit} &
\colhead{68$\%$ Range} & \colhead{Best Fit} &
\colhead{68$\%$ Range}\\
\colhead{}& \colhead{} & \colhead{(10$^{8}$ M$_{\odot}$)}&
\colhead{(10$^{8}$ M$_{\odot}$)} & \colhead{(Myr)} &
\colhead{(Myr)} & \colhead{(mag)} &
\colhead{(mag)}
}
\startdata
IRAC Undetected Stack & 4.8 & 15 & 8.6 -- 20 & 203 & 64 -- 570 & 0.03 & 0.00 -- 0.09 \\
IRAC Detected Stack & 1.2 & 92 & 51 -- 95 & 905 & 203 -- 1015 &  0.06
& 0.01 -- 0.17 
\enddata
\end{deluxetable*}
\vspace{-0.2cm}